\documentstyle[editedvolume,numreferences]{crckapb} 

\def\fileversion{v1.20a}% was \def\fileversion{v1.20}%
\def\filedate{21.6.94}%  was \def\filedate{26.1.94}%
\edef\epsfigRestoreAt{\catcode`@=\number\catcode`@\relax}%
\catcode`\@=11\relax
\ifx\undefined\@makeother                % -pks-
\def\@makeother#1{\catcode`#1=12\relax}  % -pks-
\fi                                      % -pks-
\immediate\write16{Document style option `epsfig', \fileversion\space
<\filedate> (edited by SPQR + pks)}% was <\filedate> (edited by SPQR)}%
\newcount\EPS@Height \newcount\EPS@Width \newcount\EPS@xscale
\newcount\EPS@yscale
\def\psfigdriver#1{%
  \bgroup\edef\next{\def\noexpand\tempa{#1}}%
    \uppercase\expandafter{\next}%
    \def\LN{DVITOLN03}%
    \def\DVItoPS{DVITOPS}%
    \def\DVIPS{DVIPS}%
    \def\emTeX{EMTEX}%
    \def\OzTeX{OZTEX}%
    \def\Textures{TEXTURES}%
    \global\chardef\fig@driver=0
    \ifx\tempa\LN
        \global\chardef\fig@driver=0\fi
    \ifx\tempa\DVItoPS
        \global\chardef\fig@driver=1\fi
    \ifx\tempa\DVIPS
        \global\chardef\fig@driver=2\fi
    \ifx\tempa\emTeX
        \global\chardef\fig@driver=3\fi
    \ifx\tempa\OzTeX
        \global\chardef\fig@driver=4\fi
    \ifx\tempa\Textures
        \global\chardef\fig@driver=5\fi
  \egroup
\def\psfig@start{}%
\def\psfig@end{}%
\def\epsfig@gofer{}%
\ifcase\fig@driver
\typeout{WARNING! ****
 no specials for LN03 psfig}%
\or % case 1: dvitops
\def\psfig@start{}%
\def\psfig@end{\special{dvitops: import \@p@sfilefinal \space
\@p@swidth sp \space \@p@sheight sp \space fill}%
\if@clip \typeout{Clipping not supported}\fi
\if@angle \typeout{Rotating not supported}\fi
}%
\let\epsfig@gofer\psfig@end
\or %case2 dvips
\def\psfig@start{\special{ps::[begin]  \@p@swidth \space \@p@sheight \space%
        \@p@sbbllx \space \@p@sbblly \space%
        \@p@sbburx \space \@p@sbbury \space%
        startTexFig \space }%
        \if@clip
                \if@verbose
                        \typeout{(clipped to BB) }%
                \fi
                \special{ps:: doclip \space }%
        \fi
        \if@angle              % moved after \if@clip ... \fi -pks-
                \special {ps:: \@p@sangle \space rotate \space}
        \fi
        \special{ps: plotfile \@p@sfilefinal \space }%
        \special{ps::[end] endTexFig \space }%
}%
\def\psfig@end{}%
\def\epsfig@gofer{\if@clip
                        \if@verbose
                           \typeout{(clipped to BB)}%
                        \fi
                        \epsfclipon
                  \fi
                  \epsfsetgraph{\@p@sfilefinal}%
}%
\or % case 3, emTeX
\typeout{WARNING. You must have a .bb info file with the Bounding Box
  of the pcx file}%
\def\psfig@start{}%
\def\psfig@end{\typeout{pcx import of \@p@sfilefinal}%
\if@clip \typeout{Clipping not supported}\fi
\if@angle \typeout{Rotating not supported}\fi
\raisebox{\@p@srheight sp}{\special{em: graph \@p@sfilefinal}}}%
\def\epsfig@gofer{}%
\or % case 4, OzTeX
\def\psfig@start{}%
\def\psfig@end{%
\EPS@Width\@p@swidth
\EPS@Height\@p@sheight
\divide\EPS@Width by 65781  % convert sp to bp
\divide\EPS@Height by 65781
\special{epsf=\@p@sfilefinal
\space
width=\the\EPS@Width
\space
height=\the\EPS@Height
}%
\if@clip \typeout{Clipping not supported}\fi
\if@angle \typeout{Rotating not supported}\fi
}%
\let\epsfig@gofer\psfig@end
\or % case 5, Textures
\def\psfig@end{
         \EPS@Width=\@bbw  
         \divide\EPS@Width by 1000
         \EPS@xscale=\@p@swidth \divide \EPS@xscale by \EPS@Width
         \EPS@Height=\@bbh  
         \divide\EPS@Height by 1000
         \EPS@yscale=\@p@sheight \divide \EPS@yscale by\EPS@Height
  \ifnum\EPS@xscale>\EPS@yscale\EPS@xscale=\EPS@yscale\fi
\if@clip
   \if@verbose
      \typeout{(clipped to BB)}%
   \fi
   \epsfclipon
\fi
\special{illustration \@p@sfilefinal\space scaled \the\EPS@xscale}%
}%
\def\psfig@start{}%
\let\epsfig\psfig
\else
\typeout{WARNING. *** unknown  driver - no psfig}%
\fi
}%
\newdimen\ps@dimcent
\ifx\undefined\fbox
\newdimen\fboxrule
\newdimen\fboxsep
\newdimen\ps@tempdima
\newbox\ps@tempboxa
\long\def\fbox#1{\leavevmode\setbox\ps@tempboxa\hbox{#1}\ps@tempdima\fboxrule
    \advance\ps@tempdima \fboxsep \advance\ps@tempdima \dp\ps@tempboxa
   \hbox{\lower \ps@tempdima\hbox
  {\vbox{\hrule height \fboxrule
          \hbox{\vrule width \fboxrule \hskip\fboxsep
          \vbox{\vskip\fboxsep \box\ps@tempboxa\vskip\fboxsep}\hskip
                 \fboxsep\vrule width \fboxrule}%
                 \hrule height \fboxrule}}}}%
\fi
\ifx\@ifundefined\undefined
\long\def\@ifundefined#1#2#3{\expandafter\ifx\csname
  #1\endcsname\relax#2\else#3\fi}%
\fi
\@ifundefined{typeout}%
{\gdef\typeout#1{\immediate\write\sixt@@n{#1}}}%
{\relax}%
\@ifundefined{epsfig}{}{\typeout{EPSFIG --- already loaded} }%
\@ifundefined{epsfbox}{\input epsf}{}%
\ifx\undefined\@latexerr
        \newlinechar`\^^J
        \def\@spaces{\space\space\space\space}%
        \def\@latexerr#1#2{%
        \edef\@tempc{#2}\expandafter\errhelp\expandafter{\@tempc}%
        \typeout{Error. \space see a manual for explanation.^^J
         \space\@spaces\@spaces\@spaces Type \space H <return> \space for
         immediate help.}\errmessage{#1}}%
\fi
\def\@whattodo{You tried to include a PostScript figure which
cannot be found^^JIf you press return to carry on anyway,^^J
The failed name will be printed in place of the figure.^^J
or type X to quit}%
\def\@whattodobb{You tried to include a PostScript figure which
has no^^Jbounding box, and you supplied none.^^J
If you press return to carry on anyway,^^J
The failed name will be printed in place of the figure.^^J
or type X to quit}%
\def\@nnil{\@nil}%
\def\@empty{}%
\def\@psdonoop#1\@@#2#3{}%
\def\@psdo#1:=#2\do#3{\edef\@psdotmp{#2}\ifx\@psdotmp\@empty \else
    \expandafter\@psdoloop#2,\@nil,\@nil\@@#1{#3}\fi}%
\def\@psdoloop#1,#2,#3\@@#4#5{\def#4{#1}\ifx #4\@nnil \else
       #5\def#4{#2}\ifx #4\@nnil \else#5\@ipsdoloop #3\@@#4{#5}\fi\fi}%
\def\@ipsdoloop#1,#2\@@#3#4{\def#3{#1}\ifx #3\@nnil
       \let\@nextwhile=\@psdonoop \else
      #4\relax\let\@nextwhile=\@ipsdoloop\fi\@nextwhile#2\@@#3{#4}}%
\def\@tpsdo#1:=#2\do#3{\xdef\@psdotmp{#2}\ifx\@psdotmp\@empty \else
    \@tpsdoloop#2\@nil\@nil\@@#1{#3}\fi}%
\def\@tpsdoloop#1#2\@@#3#4{\def#3{#1}\ifx #3\@nnil
       \let\@nextwhile=\@psdonoop \else
      #4\relax\let\@nextwhile=\@tpsdoloop\fi\@nextwhile#2\@@#3{#4}}%
\long\def\epsfaux#1#2:#3\\{\ifx#1\epsfpercent
   \def\testit{#2}\ifx\testit\epsfbblit
        \@atendfalse
        \epsf@atend #3 . \\%
        \if@atend
           \if@verbose
                \typeout{epsfig: found `(atend)'; continuing search}%
           \fi
        \else
                \epsfgrab #3 . . . \\%
                \epsffileokfalse\global\no@bbfalse
                \global\epsfbbfoundtrue
        \fi
   \fi\fi}%
\def\epsf@atendlit{(atend)}
\def\epsf@atend #1 #2 #3\\{%
   \def\epsf@tmp{#1}\ifx\epsf@tmp\empty
      \epsf@atend #2 #3 .\\\else
   \ifx\epsf@tmp\epsf@atendlit\@atendtrue\fi\fi}%

\chardef\trig@letter = 11
\chardef\other = 12
 
\newif\ifdebug %%% turn me on to see TeX hard at work ...
\newif\ifc@mpute %%% don't need to compute some values
\newif\if@atend
\c@mputetrue % but assume that we do
 
\let\then = \relax
\def\r@dian{pt }%
\let\r@dians = \r@dian
\let\dimensionless@nit = \r@dian
\let\dimensionless@nits = \dimensionless@nit
\def\internal@nit{sp }%
\let\internal@nits = \internal@nit
\newif\ifstillc@nverging
\def \Mess@ge #1{\ifdebug \then \message {#1} \fi}%
 
{ %%% Things that need abnormal catcodes %%%
        \catcode `\@ = \trig@letter
        \gdef \nodimen {\expandafter \n@dimen \the \dimen}%
        \gdef \term #1 #2 #3%
               {\edef \t@ {\the #1}%%% freeze parameter 1 (count, by value)
                \edef \t@@ {\expandafter \n@dimen \the #2\r@dian}%
                \t@rm {\t@} {\t@@} {#3}%
               }%
        \gdef \t@rm #1 #2 #3%
               {{%
                \count 0 = 0
                \dimen 0 = 1 \dimensionless@nit
                \dimen 2 = #2\relax
                \Mess@ge {Calculating term #1 of \nodimen 2}%
                \loop
                \ifnum  \count 0 < #1
                \then   \advance \count 0 by 1
                        \Mess@ge {Iteration \the \count 0 \space}%
                        \Multiply \dimen 0 by {\dimen 2}%
                        \Mess@ge {After multiplication, term = \nodimen 0}%
                        \Divide \dimen 0 by {\count 0}%
                        \Mess@ge {After division, term = \nodimen 0}%
                \repeat
                \Mess@ge {Final value for term #1 of
                                \nodimen 2 \space is \nodimen 0}%
                \xdef \Term {#3 = \nodimen 0 \r@dians}%
                \aftergroup \Term
               }}%
        \catcode `\p = \other
        \catcode `\t = \other
        \gdef \n@dimen #1pt{#1} %%% throw away the ``pt''
}%
 
\def \Divide #1by #2{\divide #1 by #2} %%% just a synonym
 
\def \Multiply #1by #2%%% allows division of a dimen by a dimen
       {{%%% should really freeze parameter 2 (dimen, passed by value)
        \count 0 = #1\relax
        \count 2 = #2\relax
        \count 4 = 65536
        \Mess@ge {Before scaling, count 0 = \the \count 0 \space and
                        count 2 = \the \count 2}%
        \ifnum  \count 0 > 32767 %%% do our best to avoid overflow
        \then   \divide \count 0 by 4
                \divide \count 4 by 4
        \else   \ifnum  \count 0 < -32767
                \then   \divide \count 0 by 4
                        \divide \count 4 by 4
                \else
                \fi
        \fi
        \ifnum  \count 2 > 32767 %%% while retaining reasonable accuracy
        \then   \divide \count 2 by 4
                \divide \count 4 by 4
        \else   \ifnum  \count 2 < -32767
                \then   \divide \count 2 by 4
                        \divide \count 4 by 4
                \else
                \fi
        \fi
        \multiply \count 0 by \count 2
        \divide \count 0 by \count 4
        \xdef \product {#1 = \the \count 0 \internal@nits}%
        \aftergroup \product
       }}%
 
\def\r@duce{\ifdim\dimen0 > 90\r@dian \then   % sin(x) = sin(180-x)
                \multiply\dimen0 by -1
                \advance\dimen0 by 180\r@dian
                \r@duce
            \else \ifdim\dimen0 < -90\r@dian \then  % sin(x) = sin(360+x)
                \advance\dimen0 by 360\r@dian
                \r@duce
                \fi
            \fi}%
 
\def\Sine#1%
       {{%
        \dimen 0 = #1 \r@dian
        \r@duce
        \ifdim\dimen0 = -90\r@dian \then
           \dimen4 = -1\r@dian
           \c@mputefalse
        \fi
        \ifdim\dimen0 = 90\r@dian \then
           \dimen4 = 1\r@dian
           \c@mputefalse
        \fi
        \ifdim\dimen0 = 0\r@dian \then
           \dimen4 = 0\r@dian
           \c@mputefalse
        \fi
        \ifc@mpute \then
                \divide\dimen0 by 180
                \dimen0=3.141592654\dimen0
                \dimen 2 = 3.1415926535897963\r@dian %%% a well-known constant
                \divide\dimen 2 by 2 %%% we only deal with -pi/2 : pi/2
                \Mess@ge {Sin: calculating Sin of \nodimen 0}%
                \count 0 = 1 %%% see power-series expansion for sine
                \dimen 2 = 1 \r@dian %%% ditto
                \dimen 4 = 0 \r@dian %%% ditto
                \loop
                        \ifnum  \dimen 2 = 0 %%% then we've done
                        \then   \stillc@nvergingfalse
                        \else   \stillc@nvergingtrue
                        \fi
                        \ifstillc@nverging %%% then calculate next term
                        \then   \term {\count 0} {\dimen 0} {\dimen 2}%
                                \advance \count 0 by 2
                                \count 2 = \count 0
                                \divide \count 2 by 2
                                \ifodd  \count 2 %%% signs alternate
                                \then   \advance \dimen 4 by \dimen 2
                                \else   \advance \dimen 4 by -\dimen 2
                                \fi
                \repeat
        \fi
                        \xdef \sine {\nodimen 4}%
       }}%
 
\def\Cosine#1{\ifx\sine\UnDefined\edef\Savesine{\relax}\else
                             \edef\Savesine{\sine}\fi
        {\dimen0=#1\r@dian\multiply\dimen0 by -1
         \advance\dimen0 by 90\r@dian
         \Sine{\nodimen 0}%
         \xdef\cosine{\sine}%
         \xdef\sine{\Savesine}}}
\def\psdraft{\def\@psdraft{0}}%
\def\psfull{\def\@psdraft{1}}%
\psfull
\newif\if@compress
\def\pscompress{\@compresstrue}
\def\psnocompress{\@compressfalse}
\@compressfalse
\newif\if@scalefirst
\def\psscalefirst{\@scalefirsttrue}%
\def\psrotatefirst{\@scalefirstfalse}%
\psrotatefirst
\newif\if@draftbox
\def\psnodraftbox{\@draftboxfalse}%
\@draftboxtrue
\newif\if@noisy
\@noisyfalse
\newif\ifno@bb
\newif\if@bbllx
\newif\if@bblly
\newif\if@bburx
\newif\if@bbury
\newif\if@height
\newif\if@width
\newif\if@rheight
\newif\if@rwidth
\newif\if@angle
\newif\if@clip
\newif\if@verbose
\newif\if@prologfile
\def\@p@@sprolog#1{\@prologfiletrue\def\@prologfileval{#1}}%
\def\@p@@sclip#1{\@cliptrue}%
\newif\ifepsfig@dos  % only single suffix possible
\def\epsfigdos{\epsfig@dostrue}%
\epsfig@dosfalse
\newif\ifuse@psfig
\def\ParseName#1{\expandafter\@Parse#1}%
\def\@Parse#1.#2:{\gdef\BaseName{#1}\gdef\FileType{#2}}%

\def\@p@@sfile#1{%
  \ifepsfig@dos
     \ParseName{#1:}%
  \else
     \gdef\BaseName{#1}\gdef\FileType{}%
  \fi
  \def\@p@sfile{NO FILE: #1}%
  \def\@p@sfilefinal{NO FILE: #1}%
  \openin1=#1
  \ifeof1\closein1\openin1=\BaseName.bb
    \ifeof1\closein1
      \if@bbllx                 % No postscript file but bb given explicitly.
        \if@bblly\if@bburx\if@bbury
          \def\@p@sfile{#1}%
          \def\@p@sfilefinal{#1}%
        \fi\fi\fi
      \else                     % No bounding box found.
        \@latexerr{ERROR. PostScript file #1 not found}\@whattodo
        \@p@@sbbllx{100bp}%
        \@p@@sbblly{100bp}%
        \@p@@sbburx{200bp}%
        \@p@@sbbury{200bp}%
        \psdraft
      \fi
    \else                       % Postscript file is compressed.
      \closein1%
      \edef\@p@sfile{\BaseName.bb}%
      \typeout{using BB from \@p@sfile}%
      \ifnum\fig@driver=3
        \edef\@p@sfilefinal{\BaseName.pcx}%
      \else
        \ifepsfig@dos
          \edef\@p@sfilefinal{"`gunzip -c `texfind \BaseName.{z,Z,gz}"}%
        \else
          \edef\@p@sfilefinal{"`epsfig \if@compress-c \fi#1"}%          
        \fi
      \fi
    \fi
  \else\closein1                % Postscript file is not compressed.
    \edef\@p@sfile{#1}%
    \if@compress  
      \edef\@p@sfilefinal{"`epsfig -c #1"}%
    \else
      \edef\@p@sfilefinal{#1}%
    \fi
  \fi%
}

\let\@p@@sfigure\@p@@sfile
\def\@p@@sbbllx#1{%
                                            \@bbllxtrue
                \ps@dimcent=#1
                \edef\@p@sbbllx{\number\ps@dimcent}%
                \divide\ps@dimcent by65536
                \global\edef\epsfllx{\number\ps@dimcent}%
}%
\def\@p@@sbblly#1{%
                \@bbllytrue
                \ps@dimcent=#1
                \edef\@p@sbblly{\number\ps@dimcent}%
                \divide\ps@dimcent by65536
                \global\edef\epsflly{\number\ps@dimcent}%
}%
\def\@p@@sbburx#1{%
                \@bburxtrue
                \ps@dimcent=#1
                \edef\@p@sbburx{\number\ps@dimcent}%
                \divide\ps@dimcent by65536
                \global\edef\epsfurx{\number\ps@dimcent}%
}%
\def\@p@@sbbury#1{%
                \@bburytrue
                \ps@dimcent=#1
                \edef\@p@sbbury{\number\ps@dimcent}%
                \divide\ps@dimcent by65536
                \global\edef\epsfury{\number\ps@dimcent}%
}%
\def\@p@@sheight#1{%
                \@heighttrue
                \global\epsfysize=#1
                \ps@dimcent=#1
                \edef\@p@sheight{\number\ps@dimcent}%
}%
\def\@p@@swidth#1{%
                \@widthtrue
                \global\epsfxsize=#1
                \ps@dimcent=#1
                \edef\@p@swidth{\number\ps@dimcent}% 
}%
\def\@p@@srheight#1{%
                \@rheighttrue\use@psfigtrue
                \ps@dimcent=#1
                \edef\@p@srheight{\number\ps@dimcent}%
}%
\def\@p@@srwidth#1{%
                \@rwidthtrue\use@psfigtrue
                \ps@dimcent=#1
                \edef\@p@srwidth{\number\ps@dimcent}%
}%
\def\@p@@sangle#1{%
                \use@psfigtrue
                \@angletrue
                \edef\@p@sangle{#1}%
}%
\def\@p@@ssilent#1{%
                \@verbosefalse
}%
\def\@p@@snoisy#1{%
                \@verbosetrue
}%
\def\@cs@name#1{\csname #1\endcsname}%
\def\@setparms#1=#2,{\@cs@name{@p@@s#1}{#2}}%
\def\ps@init@parms{%
                \@bbllxfalse \@bbllyfalse
                \@bburxfalse \@bburyfalse
                \@heightfalse \@widthfalse
                \@rheightfalse \@rwidthfalse
                \def\@p@sbbllx{}\def\@p@sbblly{}%
                \def\@p@sbburx{}\def\@p@sbbury{}%
                \def\@p@sheight{}\def\@p@swidth{}%
                \def\@p@srheight{}\def\@p@srwidth{}%
                \def\@p@sangle{0}%
                \def\@p@sfile{}%
                \use@psfigfalse
                \@prologfilefalse
                \def\@sc{}%
                \if@noisy
                        \@verbosetrue
                \else
                        \@verbosefalse
                \fi
                \@clipfalse
}%
\def\parse@ps@parms#1{%
                \@psdo\@psfiga:=#1\do
                   {\expandafter\@setparms\@psfiga,}%
\if@prologfile
\fi
}%
\def\bb@missing{%
        \if@verbose
            \typeout{psfig: searching \@p@sfile \space  for bounding box}%
        \fi
        \epsfgetbb{\@p@sfile}%
        \ifepsfbbfound
            \ps@dimcent=\epsfllx bp\edef\@p@sbbllx{\number\ps@dimcent}%
            \ps@dimcent=\epsflly bp\edef\@p@sbblly{\number\ps@dimcent}%
            \ps@dimcent=\epsfurx bp\edef\@p@sbburx{\number\ps@dimcent}%
            \ps@dimcent=\epsfury bp\edef\@p@sbbury{\number\ps@dimcent}%
        \else
            \epsfbbfoundfalse
        \fi
}
\newdimen\p@intvaluex
\newdimen\p@intvaluey
\def\rotate@#1#2{{\dimen0=#1 sp\dimen1=#2 sp
                  \global\p@intvaluex=\cosine\dimen0
                  \dimen3=\sine\dimen1
                  \global\advance\p@intvaluex by -\dimen3
                  \global\p@intvaluey=\sine\dimen0
                  \dimen3=\cosine\dimen1
                  \global\advance\p@intvaluey by \dimen3
                  }}%
\def\compute@bb{%
                \epsfbbfoundfalse
                \if@bbllx\epsfbbfoundtrue\fi
                \if@bblly\epsfbbfoundtrue\fi
                \if@bburx\epsfbbfoundtrue\fi
                \if@bbury\epsfbbfoundtrue\fi
                \ifepsfbbfound\else\bb@missing\fi
                \ifepsfbbfound\else
                \@latexerr{ERROR. cannot locate BoundingBox}\@whattodobb
                        \@p@@sbbllx{100bp}%
                        \@p@@sbblly{100bp}%
                        \@p@@sbburx{200bp}%
                        \@p@@sbbury{200bp}%
                        \no@bbtrue
                        \psdraft
                \fi
                \count203=\@p@sbburx
                \count204=\@p@sbbury
                \advance\count203 by -\@p@sbbllx
                \advance\count204 by -\@p@sbblly
                \edef\ps@bbw{\number\count203}%
                \edef\ps@bbh{\number\count204}%
                 \edef\@bbw{\number\count203}%
                \edef\@bbh{\number\count204}%
               \if@angle
                        \Sine{\@p@sangle}\Cosine{\@p@sangle}%
 
{\ps@dimcent=\maxdimen\xdef\r@p@sbbllx{\number\ps@dimcent}%
 
\xdef\r@p@sbblly{\number\ps@dimcent}%
 
\xdef\r@p@sbburx{-\number\ps@dimcent}%
 
\xdef\r@p@sbbury{-\number\ps@dimcent}}%
                        \def\minmaxtest{%
                           \ifnum\number\p@intvaluex<\r@p@sbbllx
                              \xdef\r@p@sbbllx{\number\p@intvaluex}\fi
                           \ifnum\number\p@intvaluex>\r@p@sbburx
                              \xdef\r@p@sbburx{\number\p@intvaluex}\fi
                           \ifnum\number\p@intvaluey<\r@p@sbblly
                              \xdef\r@p@sbblly{\number\p@intvaluey}\fi
                           \ifnum\number\p@intvaluey>\r@p@sbbury
                              \xdef\r@p@sbbury{\number\p@intvaluey}\fi
                           }%
                        \rotate@{\@p@sbbllx}{\@p@sbblly}%
                        \minmaxtest
                        \rotate@{\@p@sbbllx}{\@p@sbbury}%
                        \minmaxtest
                        \rotate@{\@p@sbburx}{\@p@sbblly}%
                        \minmaxtest
                        \rotate@{\@p@sbburx}{\@p@sbbury}%
                        \minmaxtest
 
\edef\@p@sbbllx{\r@p@sbbllx}\edef\@p@sbblly{\r@p@sbblly}%
 
\edef\@p@sbburx{\r@p@sbburx}\edef\@p@sbbury{\r@p@sbbury}%
                \fi
                \count203=\@p@sbburx
                \count204=\@p@sbbury
                \advance\count203 by -\@p@sbbllx
                \advance\count204 by -\@p@sbblly
                \edef\@bbw{\number\count203}%
                \edef\@bbh{\number\count204}%
}%
\def\in@hundreds#1#2#3{\count240=#2 \count241=#3
                     \count100=\count240        % 100 is first digit #2/#3
                     \divide\count100 by \count241
                     \count101=\count100
                     \multiply\count101 by \count241
                     \advance\count240 by -\count101
                     \multiply\count240 by 10
                     \count101=\count240        %101 is second digit of #2/#3
                     \divide\count101 by \count241
                     \count102=\count101
                     \multiply\count102 by \count241
                     \advance\count240 by -\count102
                     \multiply\count240 by 10
                     \count102=\count240        % 102 is the third digit
                     \divide\count102 by \count241
                     \count200=#1\count205=0
                     \count201=\count200
                        \multiply\count201 by \count100
                        \advance\count205 by \count201
                     \count201=\count200
                        \divide\count201 by 10
                        \multiply\count201 by \count101
                        \advance\count205 by \count201
                     \count201=\count200
                        \divide\count201 by 100
                        \multiply\count201 by \count102
                        \advance\count205 by \count201
                     \edef\@result{\number\count205}%
}%
\def\compute@wfromh{%
                \in@hundreds{\@p@sheight}{\@bbw}{\@bbh}%
                \edef\@p@swidth{\@result}%
}%
\def\compute@hfromw{%
                \in@hundreds{\@p@swidth}{\@bbh}{\@bbw}%
                \edef\@p@sheight{\@result}%
}%
\def\compute@handw{%
                \if@height
                        \if@width
                        \else
                                \compute@wfromh
                        \fi
                \else
                        \if@width
                                \compute@hfromw
                        \else
                                \edef\@p@sheight{\@bbh}%
                                \edef\@p@swidth{\@bbw}%
                        \fi
                \fi
}%
\def\compute@resv{%
                \if@rheight \else \edef\@p@srheight{\@p@sheight} \fi
                \if@rwidth \else \edef\@p@srwidth{\@p@swidth} \fi
}%
\def\compute@sizes{%
        \if@scalefirst\if@angle
        \if@width
           \in@hundreds{\@p@swidth}{\@bbw}{\ps@bbw}%
           \edef\@p@swidth{\@result}%
        \fi
        \if@height
           \in@hundreds{\@p@sheight}{\@bbh}{\ps@bbh}%
           \edef\@p@sheight{\@result}%
        \fi
        \fi\fi
        \compute@handw
        \compute@resv
}
\long\def\graphic@verb#1{\def\next{#1}%
  {\expandafter\graphic@strip\meaning\next}}
\def\graphic@strip#1>{}
\def\graphic@zapspace#1{%
  #1\ifx\graphic@zapspace#1\graphic@zapspace%
  \else\expandafter\graphic@zapspace%
  \fi}
\def\psfig#1{%
\edef\@tempa{\graphic@zapspace#1{}}%
\ifvmode\leavevmode\fi\vbox {%
        \ps@init@parms
        \parse@ps@parms{\@tempa}%
        \ifnum\@psdraft=1
                \typeout{[\@p@sfilefinal]}%
                \if@verbose
                        \typeout{epsfig: using PSFIG macros}%
                \fi
                \psfig@method
        \else
                \epsfig@draft
        \fi
}
}%
\def\graphic@zapspace#1{%
  #1\ifx\graphic@zapspace#1\graphic@zapspace%
  \else\expandafter\graphic@zapspace%
  \fi}
\def\epsfig#1{%
\edef\@tempa{\graphic@zapspace#1{}}%
\ifvmode\leavevmode\fi\vbox {%
        \ps@init@parms
        \parse@ps@parms{\@tempa}%
        \ifnum\@psdraft=1
          \if@angle\use@psfigtrue\fi
          {\ifnum\fig@driver=1\global\use@psfigtrue\fi}%
          {\ifnum\fig@driver=3\global\use@psfigtrue\fi}%
          {\ifnum\fig@driver=4\global\use@psfigtrue\fi}%
          {\ifnum\fig@driver=5\global\use@psfigtrue\fi}%
                \ifuse@psfig
                        \if@verbose
                                \typeout{epsfig: using PSFIG macros}%
                        \fi
                        \psfig@method
                \else
                        \if@verbose
                                \typeout{epsfig: using EPSF macros}%
                        \fi
                        \epsf@method
                \fi
        \else
                \epsfig@draft
        \fi
}%
}%

\def\epsf@method{%
        \epsfbbfoundfalse
        \if@bbllx\epsfbbfoundtrue\fi
        \if@bblly\epsfbbfoundtrue\fi
        \if@bburx\epsfbbfoundtrue\fi
        \if@bbury\epsfbbfoundtrue\fi
        \ifepsfbbfound\else\epsfgetbb{\@p@sfile}\fi
        \ifepsfbbfound
           \typeout{<\@p@sfilefinal>}%
           \epsfig@gofer
        \else
          \@latexerr{ERROR - Cannot locate BoundingBox}\@whattodobb
          \@p@@sbbllx{100bp}%
          \@p@@sbblly{100bp}%
          \@p@@sbburx{200bp}%
          \@p@@sbbury{200bp}%
                \count203=\@p@sbburx
                \count204=\@p@sbbury
                \advance\count203 by -\@p@sbbllx
                \advance\count204 by -\@p@sbblly
                \edef\@bbw{\number\count203}%
                \edef\@bbh{\number\count204}%
          \compute@sizes
          \epsfig@@draft
       \fi
}%
\def\psfig@method{%
        \compute@bb
        \ifepsfbbfound
          \compute@sizes
          \psfig@start
          \vbox to \@p@srheight sp{\hbox to \@p@srwidth 
            sp{\hss}\vss\psfig@end}%
        \else
           \epsfig@draft
        \fi
}%
\def\epsfig@draft{\compute@bb\compute@sizes\epsfig@@draft}%
\def\epsfig@@draft{%
\typeout{<(draft only) \@p@sfilefinal>}%
\if@draftbox
        \hbox{{\fboxsep0pt\fbox{\vbox to \@p@srheight sp{%
        \vss\hbox to \@p@srwidth sp{ \hss 
           \expandafter\Literally\@p@sfilefinal\@nil
                          \hss }\vss
        }}}}%
\else
        \vbox to \@p@srheight sp{%
        \vss\hbox to \@p@srwidth sp{\hss}\vss}%
\fi
}%
\def\Literally#1\@nil{{\tt\graphic@verb{#1}}}
\psfigdriver{dvips}%
\epsfigRestoreAt

\begin{opening}
\title{M-Theory and N=2 Strings}
\author{Emil Martinec}
\institute{Enrico Fermi Institute and Dept. of Physics\\
           5640 South Ellis Ave., Chicago, IL 60637-1433}
\end{opening}
\begin{document}

%%%%%%%%%%%%%%%%%%%%%%%%%%%%%%%%%%%%%%%%%%%%%%%%%%%%%%%%%

\def\pref#1{(\ref{#1})}

\def\ie{{i.e.}}
\def\eg{{e.g.}}
\def\cf{{c.f.}}
\def\etal{{et.al.}}
\def\etc{{etc.}}

\def\inbar{\,\vrule height1.5ex width.4pt depth0pt}
\def\IR{\relax{\rm I\kern-.18em R}}
\def\IC{\relax\hbox{$\inbar\kern-.3em{\rm C}$}}
\def\IH{\relax{\rm I\kern-.18em H}}
\def\IO{\relax\hbox{$\inbar\kern-.3em{\rm O}$}}
\def\IK{\relax{\rm I\kern-.18em K}}
\def\IP{\relax{\rm I\kern-.18em P}}
\def\Z{{\bf Z}}
\def\Pone{{\IC\rm P^1}}
\def\One{{1\hskip -3pt {\rm l}}}

\def\beq{\begin{equation}}
\def\eeq{\end{equation}}

\def\sst{\scriptscriptstyle}
\def\tst#1{{\textstyle #1}}
\def\frac#1#2{{#1\over#2}}
\def\coeff#1#2{{\textstyle{#1\over #2}}}
\def\half{\frac12}
\def\hf{{\textstyle\half}}

%%%%%%%%%%%%%%%%%%%%%%%%%%%%%%%%%%%%%%%%%%%%%%%%%%%%%%%%%%%%%%
\def\sdtimes{\mathbin{\hbox{\hskip2pt\vrule height 4.1pt depth -.3pt width
.25pt \hskip-2pt$\times$}}}
\def\bra#1{\left\langle #1\right|}
\def\ket#1{\left| #1\right\rangle}
\def\vev#1{\left\langle #1 \right\rangle}
\def\det{{\rm det}}
\def\tr{{\rm tr}}
\def\mod{{\rm mod}}
\def\sinh{{\rm sinh}}
\def\cosh{{\rm cosh}}
\def\sgn{{\rm sgn}}
\def\det{{\rm det}}
\def\exp{{\rm exp}}
\def\sh{{\rm sh}}
\def\ch{{\rm ch}}
\def\d{{\partial}}
\def\dbar{{\bar\partial}}

%%%%%%%%%%%%%%%%%%%%%%%%%%%%%%%%%%%%%%%%%%%%%%%%%%%%%%%%%%%%%%
\def\np{{\it Nucl. Phys. }}
\def\npb{{\it Nucl. Phys. }}
\def\pl{{\it Phys. Lett. }}
\def\plb{{\it Phys. Lett. }}
\def\pr{{\it Phys. Rev. }}
\def\prd{{\it Phys. Rev. }}
\def\ap{{\it Ann. Phys., NY }}
\def\prl{{\it Phys. Rev. Lett. }}
\def\mpl{{\it Mod. Phys. Lett. }}
\def\cmp{{\it Comm. Math. Phys. }}
\def\grg{{\it Gen. Rel. and Grav. }}
\def\cqg{{\it Class. Quant. Grav. }}
\def\ijmp{{\it Int. J. Mod. Phys. }}
\def\jmp{{\it J. Math. Phys. }}
\def\nextline{\hfil\break}
\catcode`\@=11
\def\slash#1{\mathord{\mathpalette\c@ncel{#1}}}
\overfullrule=0pt
\def\AA{{\cal A}}
\def\BB{{\cal B}}
\def\CC{{\cal C}}
\def\DD{{\cal D}}
\def\EE{{\cal E}}
\def\FF{{\cal F}}
\def\GG{{\cal G}}
\def\HH{{\cal H}}
\def\II{{\cal I}}
\def\JJ{{\cal J}}
\def\KK{{\cal K}}
\def\LL{{\cal L}}
\def\MM{{\cal M}}
\def\NN{{\cal N}}
\def\OO{{\cal O}}
\def\PP{{\cal P}}
\def\QQ{{\cal Q}}
\def\RR{{\cal R}}
\def\SS{{\cal S}}
\def\TT{{\cal T}}
\def\UU{{\cal U}}
\def\VV{{\cal V}}
\def\WW{{\cal W}}
\def\XX{{\cal X}}
\def\YY{{\cal Y}}
\def\ZZ{{\cal Z}}
\def\lam{\lambda}
\def\eps{\epsilon}
\def\vareps{\varepsilon}
\def\underrel#1\over#2{\mathrel{\mathop{\kern\z@#1}\limits_{#2}}}
\def\lapprox{{\underrel{\scriptstyle<}\over\sim}}
\def\lessapprox{{\buildrel{<}\over{\scriptstyle\sim}}}
\catcode`\@=12
%%%%%%%%%%%%%%%%%%%%%%%%%%%%%%%%%%%%%%%%%%%%%%%%%%%%%%%%%%%%%%
% macros specific to this article
\def\sdiff{{{\sl SDiff}_2}}
\def\ibar{{\bar i}}
\def\jbar{{\bar j}}
\def\xbar{{\bar x}}
\def\zbar{{\bar z}}
\def\tbar{{\bar\theta}}
\def\thetabar{{\bar\theta}}
\def\psibar{{\bar\psi}}
\def\phibar{{\bar\phi}}
\def\abar{{\bar a}}
\def\bbar{{\bar b}}
\def\kbar{{\bar k}}
\def\mbar{{\bar m}}
\def\vv{{\bf v}}
\def\s{{\bf S}}
\def\su{{\rm SU}}
\def\ij{{i\bar j}}
\def\kahler{{K\"ahler}}
\def\lstr{\ell_{\rm str}}
\def\lpl{\ell_{\rm pl}}
\def\gstr{g_{\rm str}}
\def\gym{g_{\sst\rm YM}}
%%%%%%%%%%%%%%%%%%%%%%%%%%%%%%%%%%%%%%%%%%%%%%%%%%%%%%%%%%%%%%
%%%%%%%%%%%%%%%%%%%%%%%%%%%%%%%%%%%%%%%%%%%%%%%%%%%%%%%%%%%%%%

\begin{abstract}
N=2 heterotic strings may provide a window into the physics of
M-theory radically different than that found via the other
supersymmetric string theories.  
In addition to their supersymmetric structure, these strings
carry a four-dimensional self-dual structure, and appear
to be completely integrable systems with a stringy density of states.
These lectures give an overview of N=2 heterotic strings,
as well as a brief discussion of possible applications of
both ordinary and heterotic N=2 strings to D-branes and matrix
theory.
\end{abstract}

\section{Introduction}

\index{heterotic string}
A few years ago, if asked to describe string theory,
the average practitioner would have classified its different
manifestations according to their various worldsheet gauge principles.
On the 1+1 dimensional worldsheet, there can be $(p,q)$ supersymmetries
that square to translations along the (left,right)-handed light cone; 
one says that the worldsheet has $(p,q)$ gauged supersymmetry.
The bosonic string has no supersymmetry; $p=q=0$.
The supersymmetric string theories have, say, $q=1$.
Thus type IIA/B string theory has (1,1) supersymmetry.  The type I/IA
strings are the orbifold of these by worldsheet parity,
and the heterotic strings are in the class (0,1).
Remarkably, we now understand that all the supersymmetric string theories
-- type IIA/B, type I, and heterotic -- 
appear to describe asymptotic expansions of a
single nonperturbative master theory: M-theory.   
This theory has many miraculous duality properties that are now only
beginning to be unravelled; other lecturers at this school will
review the current state of affairs.  In these lectures, I will
give an overview of a relatively unexplored corner of string theory,
namely the N=2 strings \cite{ovone,ovtwo,marcusrev}
(more specifically strings with (2,2) or (2,1) 
gauged worldsheet supersymmetry).

The driving force behind the recent unification has been the recognition
of the fundamental importance of the spacetime
(as opposed to worldsheet) supersymmetry algebra.
The small (BPS) representations of the supersymmetry algebra
form a quasi-topological sector of the theory.
By tracking this BPS spectrum across moduli space, one can deduce
the interconnections of the various string limits.  The issue
we wish to address is the role of the heterotic (2,1) string, which is
also a stringy realization of spacetime supersymmetry and
therefore ought to play a role somehow.  As we will see
\cite{kmone,kmtwo},
many of the basic objects of M-theory are realized in the (2,1) string.
Self-duality \index{self-duality} and integrability are further
features of the (2,1) string, arising
from the chiral sector with N=2 worldsheet supersymmetry.

We will see that the chiral critical dimensions of the 
(2,1) string are $d=4$ (2 space, 2 time) for the N=2 sector,
and $d=12$ (10 space, 2 time) for the N=1 sector.
To see the relation to M-theory, consider standard heterotic target space
geometry.  Here this is a 2+2 dimensional base manifold 
for the dimensions common to both chiralities, with 
the additional left-movers parametrizing an eight-dimensional
torus of stringy dimensions
fibered over it (see figure 1).  

\vskip 1cm
{\vbox{{\epsfxsize=2.5in
        \nopagebreak[3]
    \centerline{\epsfbox{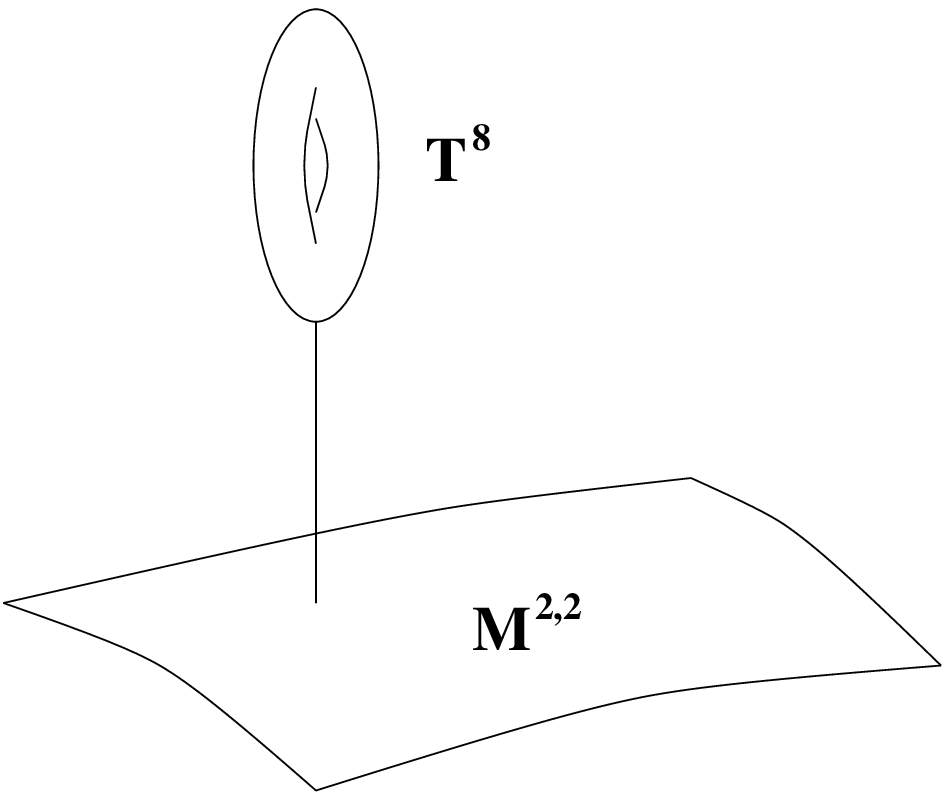}}
        \nopagebreak[3]
    {\raggedright\it \vbox{
{\bf Figure 1.}
{\it (2,1) heterotic geometry.  }
 }}}}
    \bigskip}

The geometrical fields are
the graviton $h$, antisymmetric tensor field $b$, gravitino $\chi$, and the
gauge connection $\vareps$ on the fiber.  These fields are further restricted
by the extra constraints of N=2 local supersymmetry on the worldsheet,
giving rise to prepotentials
\begin{eqnarray}
  h_{\mu\nu}+b_{\mu\nu}	&\rightarrow& I_\mu^\lam\d_\lam a_\nu\nonumber\\
  \vareps_\mu^a 	&\rightarrow& I_\mu^\lam\d_\lam\varphi^a\nonumber\\
  \chi_\mu^\alpha 	&\rightarrow& I_\mu^\lam\d_\lam\psi^\alpha\ .
\label{prepotl}
\end{eqnarray}
Here $I_\mu^\lam$ is an almost complex structure on the base space,
which is required by the local N=2 worldsheet supersymmetry.
Under linearized gauge transformations, these restricted fields
transform as
\begin{eqnarray}
 \delta(h+b)=\d\xi 	&\rightarrow& \delta a\sim \xi\nonumber\\
 \delta\vareps=\d\Lambda&\rightarrow& \delta\varphi\sim\Lambda\nonumber\\
 \delta\chi=\d\eta	&\rightarrow& \delta\psi\sim\eta\ .
\end{eqnarray}
In other words, the remnant fields are Nambu-Goldstone fields
of spontaneously broken symmetries (spacetime antisymmetric
tensor field gauge transformations, translations, supersymmetries)
on a D-brane. 
\index{D-brane} 
More precisely, one has in a complex coordinate basis
$\vareps=i(\d-\dbar)\varphi+(\d+\dbar)\theta=\d(\theta+i\varphi)
+\dbar(\theta-i\varphi)$, where the (real) gauge symmetry is
$\delta\theta=\Lambda$.  However, one can go to a holomorphic 
basis by complexifying the gauge group $G$; then $\varphi$ is a coordinate
on $G_{\bf C}/G$ whose dynamics is determined by holomorphic gauge
symmetry much as in the 2d WZW model \cite{nairschiff,LMNS,hull}.
This virtually guarantees us a connection to brane physics,
since brane dynamics is almost by definition given by the nonlinear
Lagrangian of the spacetime symmetries broken by the brane.
A picture of this aspect of (2,1) string theory is shown
in figure 2.  The (2,1) string worldsheet maps into
the worldvolume of a D-brane, which is itself embedded in
spacetime.  The quanta of this brane are the (2,1)
strings themselves.  Since the transverse fluctuations of
the brane can be traced to those of the fiber connection 
$\vareps\sim\d\varphi$, while the longitudinal directions
are those of the base space, we see that spacetime itself
is the total space of the heterotic geometry of figure 1.

\vskip .5cm
{\vbox{{\epsfxsize=5in
        \nopagebreak[3]
    \centerline{\epsfbox{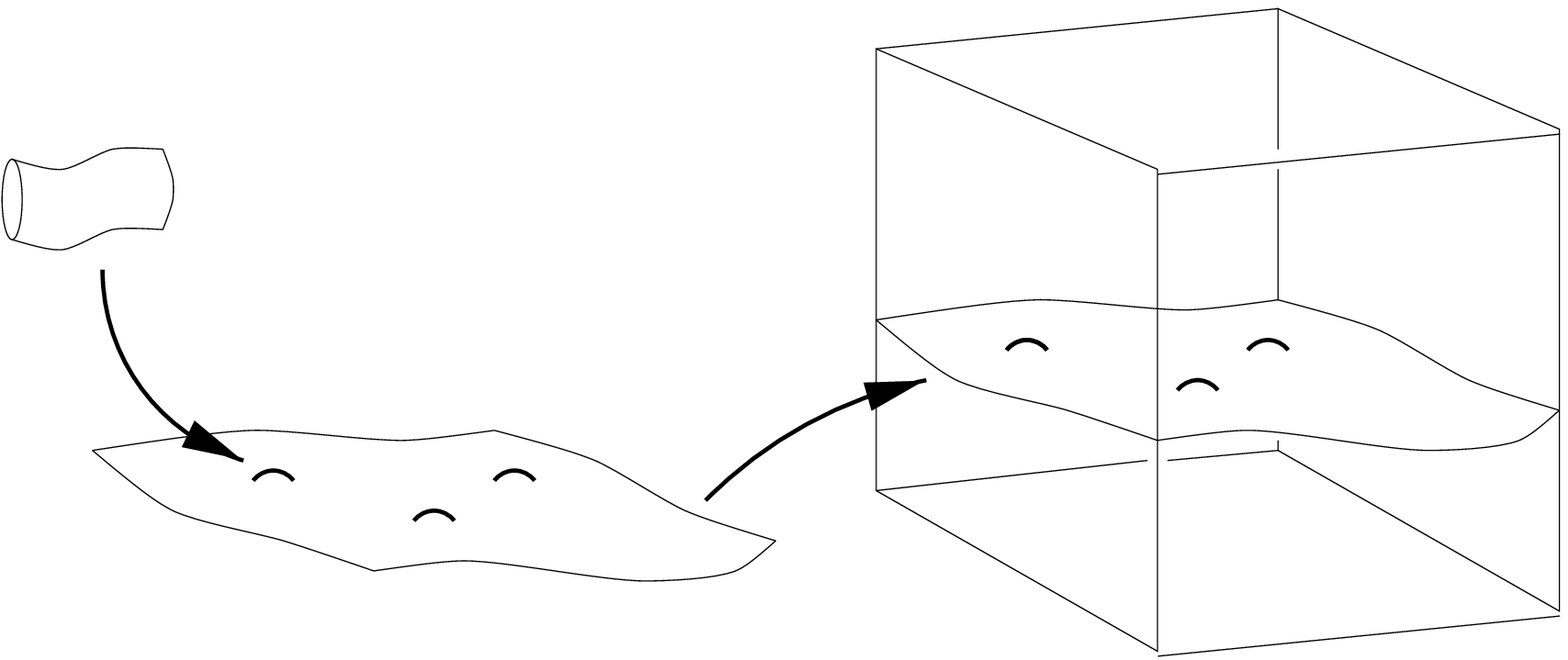}}
        \nopagebreak[3]
    {\raggedright\it \vbox{
{\bf Figure 2.}
{\it Chain of brane embeddings implied by (2,1) string states.}
 }}}}
    \bigskip}

The dimension of the brane is determined by the number of independent 
momentum components in string vertex operators $\OO_\alpha e^{ik\cdot x}$.
When the (2,1) string target space is $\IR^{2,2}\times T^8$,
this kinematics is 1+1 or 2+1 dimensional; when the spatial
dimensions are further compactified, the kinematics is more
or less 9+1 dimensional -- the target is a kind of ninebrane.

As mentioned above, an additional geometric structure is self-duality.
\index{self-duality}
The almost complex structure $I_\mu^\lam$ is one of a triplet
of such structures preserved by the target space geometry
(these almost complex structures are not integrable,
since $(\d-\dbar)I=db=\sl torsion$).
Thus one can bring to bear the machinery of twistor theory to characterize
the classical solution space.  One of the characteristic
properties of self-dual gravity is the symmetry group of 
area preserving diffeomorphisms $\sdiff\sim SU(\infty)$,
suggesting a connection to matrix theory \cite{bfss}.

There are indeed a number of intriguing analogies among the matrix model
of M-theory, (2,1) strings, and other matrix models:
\index{matrix model}
\begin{enumerate}
  \item The (BFSS) matrix model of M-theory \cite{bfss} realizes the
	membrane as a collective phenomenon of D0-brane supergravitons,
	\ie\ as a state in the large N collective field theory.
	In a sense, the BFSS matrix model gives a map from a
	{\it single} noncommutative torus into `spacetime'.  In the
	graviton limit, the matrices approximately commute; in the membrane
	limit, the commutators are large.
	Similarly, the (2,1) string describes a map from a {\it single}
	brane into spacetime.  
  \item Other examples where large N collective field theory
 	generates string theory as an asymptotic expansion around
	a particular master field include:
\begin{enumerate}
\item The 1+1d noncritical string based on the original matrix 
	model of \cite{bpiz}.
\item 2d Yang-Mills \cite{twodYM}.
\item 2+1d SU(k) Chern-Simons theory as $k,N\rightarrow\infty$ \cite{douglas}.
\item 2+2d self-dual gravity \cite{husward},\cite{ovone}.
	This connection will be outlined in section 2 below.
\end{enumerate}
\item (2,1) string perturbation theory is an expansion
around the infinite tension limit; 
$T_{\sst \rm brane}\sim g_{\sst\rm  str}^{-2}$.  In other words,
it is an expansion in low energy relative to the brane tension.
This is similar to 1+1d noncritical string theory, where
the effective expansion parameter is the energy of excitations
relative to some scale (in that case the cosmological constant $\mu$
of Liouville theory, or equivalently the fermi energy of
the matrix partons).
\end{enumerate}
Thus one sees common themes cropping up in diverse settings. 

The focus of these lectures will be the structure of N=2 strings,
and their possible relation to matrix dynamics.  We begin
in section 2 with the somewhat simpler (2,2) string,
in order to introduce the novel features of local N=2 worldsheet
supersymmetry and its associated self-duality structure.
Included are an overview of the relation between the self-dual
gravity of the (2,2) string and $\sdiff$ dynamics;
as well as an aside on the D-brane spectrum of (2,2) strings,
which leads to yet another interesting connection to matrix theory.
Section 3 introduces (2,1) strings -- their worldsheet gauge algebra,
spectrum, and simple compactifications.  The target space effective
action for (2,1) strings is derived in section 4; 
parallels with matrix theory are explored in section 5.

\section{(2,2) Strings}

The gauged N=2 supersymmetry on the string worldsheet consists
of currents $T$ (the stress tensor), $G^\pm$ (the two supercurrents), 
and $J$ (a U(1) R-symmetry current).  Their algebra is schematically
\begin{eqnarray}
[T,G^{\pm}]&\sim & \coeff32 G^\pm\nonumber\\
{[T,J]} &\sim & J\nonumber\\
{[G^+,G^-]}&\sim & T+J\label{Ntwoalg}\\
{[J,G^{\pm}]}&\sim & \pm G^\pm\ .\nonumber
\end{eqnarray}
In the conformal gauge, the contributions to the Virasoro 
central charge from the Faddeev-Popov ghosts is
$-26+2\cdot 11-2=-6$, so that the critical dimension is
$d_{\rm crit}=2/3\;c_{\rm matter}=4$.
A free field representation of the above currents is
\begin{eqnarray}
  T&=& -\hf\d X\cdot\d X-\psi\d\psi\nonumber\\
  J&=& I_{\mu\nu}\psi^\mu\psi^\nu\qquad,\qquad\qquad \mu=0,...,3\nonumber\\
  G^\pm&=& (\eta_{\mu\nu}\pm I_{\mu\nu})\psi^\mu\d X^\nu 
	= G_{\sst\rm N=1}\pm G_2\ .
\end{eqnarray}
Here, $I_{\mu\nu}$ is a self-dual tensor which acts as a complex
structure.  Preserving this structure under analytic continuation
requires the signature to be Euclidean (4+0) or ultrahyperbolic (2+2).
Even though the indefinite signature case has more than one negative
metric string coordinate, there are no negative metric states.
Each gauge invariance removes two fields' worth of degrees of freedom;
in fact, the two bosonic ($T,J$) and two fermionic ($G^\pm$) 
gauge symmetries kill {\it all} the oscillator modes
of the string.

One of the powerful consequences of this fact is the triviality
of the string S-matrix.  On the one hand, there can
be no Regge behavior in scattering amplitudes, since the 
sequence of poles in the S-matrix are associated with
physical oscillator excitations of the string, of which there
are none.  On the other hand, the covariant formalism generates
a Koba-Nielsen (integral) representation for the S-matrix
amplitude, which exhibits such poles.  
The tension between these two properties is resolved
by the vanishing of any amplitude beyond the three-point function
(the three-point function is protected because it does not involve
Koba-Nielsen integrals).  The single nontrivial S-matrix 
element is\footnote{Similar to the 1+1d noncritical string, where the
`bulk' S-matrix is trivial, the effective action is cubic (see below).}
\beq
  \vev{V(1)V(2)V(3)}=(k_1\cdot I\cdot k_2)_\ell(k_1\cdot I\cdot k_2)_r\ .
\label{Stwotwo}
\eeq
Since the only center of mass of the string can be excited,
it is possible to find a representative of the string vertex operator
which is a simple exponential 
\beq
  V=\bar\Sigma_{gh}\Sigma_{gh}\; e^{ik\cdot x}
\label{Vtwotwo}
\eeq
apart from ghost (measure) factors $\Sigma_{gh}$ (see Appendix A
for a brief summary).  A BRST-equivalent representative (or `picture')
of the vertex represents it as an integral over (2,2) superspace,
\beq
  V=\int d^2\theta d^2\bar\theta\;e^{ik\cdot X} =
	\int d\theta d\bar\theta|_{\sst\rm N=1}
		(k^\mu I_{\mu\nu}\bar D_{\sst\rm N=1}X^\nu)
		(k^\lam I_{\lam\sigma}D_{\sst\rm N=1}X^\sigma)
			e^{ik\cdot X}\ ,
\eeq
\ie\ from the N=1 point of view the single physical state
is a graviton whose only physical polarization is a fluctuation
in the Kahler potential --
$g_{i\jbar}=\d_i\d_\jbar K$ in complex coordinates.

Another important feature of the (2,2) string is the absence
of target space supersymmetry.  Consider a Wilson line of the 
U(1) R-current,
$exp[i\alpha\int^{\bar z}J]$.  Parallel transport
of a complex fermion $\psi$ around the point $z$ picks up a phase
$exp[i\alpha]$; however $J$ is gauged, and so the phase 
cannot have physical significance.  In particular, there is no
physical distinction between periodic (NS) and antiperiodic (R)
boundary conditions, hence no target space fermions and
no target space supersymmetry.

\index{effective action}
The effective action which generates the S-matrix \pref{Stwotwo} is
\beq
  S_{\rm eff}=\frac1{\gstr^2}\int d^4x[\hf\d K\cdot\dbar K+
	\coeff13 K\d\dbar K\wedge\d\dbar K]\ ,
\label{twotwoact}
\eeq
from which one obtains the equation of motion
\beq
  I\wedge\d\dbar K+\d\dbar K\wedge\d\dbar K=0\ ,
\label{pleb}
\eeq
known as the Plebanski equation.  This is the equation that 
governs the dynamics of self-dual gravity.  
It is a straighforward exercise in 2+2 kinematics 
to show that the quartic S-matrix computed from the action
\pref{twotwoact} vanishes \cite{ovone}.

An alternate route
to the Plebanski equation proceeds from the generalized
beta-function of the background sigma model for string propagation.
In (2,2) supersymmetry, there are two kinds of scalar superfields:
chiral and twisted chiral \cite{ghr}.  The simplest situation
has the sigma model background described entirely in terms of 
chiral superfields; then the target space holonomy lies in U(d).
The sigma model beta function equations 
\index{sigma model}
are the conditions for SU(d) holonomy
\beq
  R_{\mu\nu}[\Gamma]=2\nabla_\mu\nabla_\nu\Phi
\label{betafneq}
\eeq
($\Gamma^{\mu}_{\nu\lam}=\{{\mu\atop\nu\lam}\}-\half H^\mu_{\nu\lam}$
is the connection with torsion $H$).  
These equations may be integrated
in complex coordinates to yield
\beq
  {\sl log}{\rm det}[g_{\ij}]=2\Phi+f(x)+\bar f(\xbar)\ .
\label{sdg}
\eeq
A further analysis of the 
conditions for (2,2) supersymmetry
in the sigma model coupled to worldsheet gravity \cite{kounnas}
shows that the dilaton $\Phi$ is locally the real part of a holomorphic
function; thus {\it locally} one can choose coordinates such that
the dynamical equation is
\beq
  \det[g_{\ij}]=1\ .
\label{gfixed}
\eeq
Some solutions of these equations have been described in \cite{kounnas}.
The single physical degree of freedom of the N=2 string is the center
of mass mode $K(x,\xbar)$ describing fluctuations of the \kahler\ 
potential $g_{\ij}=I_\ij+\d_i\d_\jbar K$,  
where $I$ is the background \kahler\ form.
Expanding \pref{gfixed} in this way yields the Plebanski
equation \pref{pleb}.
Note that although $K$ contains the field-theoretic degrees of freedom
of the string, there may be additional moduli in the global
modes of the metric, antisymmetric tensor, and dilaton -- for instance
the action \pref{twotwoact} on K3 depends implicitly on the full 80 moduli
of the conformal field theory, as well as the string coupling 
$\kappa= e^{-2\Phi}$.  These are analogues of the special states
of 2d noncritical string theory \cite{polyakov}
that exist only for discrete momenta.

Backgrounds involving one chiral and one twisted chiral superfield
must have two commuting complex structures \cite{ghr}
and are thus essentially trivial \cite{hulltwist}
-- the quasi-\kahler\ potential $\tilde K$ describing the background
geometry satisfies a free field equation.
When spacetime has a translational
Killing vector field with compact orbits, these backgrounds
are related to the above self-dual geometry by
T-duality \cite{ghr,kounnas}.
Thus one might call the theory discribed by \pref{twotwoact}
the N=2B theory, and the theory described by the free quasi-\kahler\
potential $\tilde K$ the N=2A string.

There are also interesting solutions which fall outside the class
described by constant dilaton, for example the NS instanton 
\cite{axinst},\cite{kounnas}
\begin{eqnarray}
  e^{2\Phi}&=&e^{2\Phi_0}+\frac{n\alpha'}{r^2}\nonumber\\
  H_{\mu\nu\lam}&=&-{\eps_{\mu\nu\lam}}^\rho\nabla_\rho\Phi\nonumber\\
  G_{\mu\nu}&=&e^{2\Phi}\delta_{\mu\nu}\ ,
\label{NSinst}
\end{eqnarray}
which is the counterpart of the NS five-brane in the usual 10d
theory (an instanton is the magnetic dual to a string in 4d).
In particular, this object is not that same as the D-instanton
encountered below.

The Ricci-flat four dimensional geometry required by the
N=2 string admits a hyperKahler structure, and the metric is
self-dual.  
\index{self-duality}
It is easy to see that the N=2 local supersymmetry
generates N=4 global supersymmetry in the critical dimension 
$d_{\rm crit}=4$.  The canonical normalization of the U(1) R-current
\beq
  J(z)J(w)\sim\frac1{(z-w)^2}
\eeq
implies that the exponential fields $J^\pm(z)=exp[\pm i\sqrt2\int^z J]$,
together with $J(z)$, form an SU(2) current algebra;
furthermore, $J^\mp G^\pm\sim(z-w)^{-1}\tilde G^\pm$ are two additional
supersymmetry currents.  These enlarge the N=2 currents
\pref{Ntwoalg} to the N=4 current multiplet.
Since any solution to the background field equations of motion
preserves an SU(2) symmetry in the tangent space
(in Euclidean signature; SL(2,R) in ultrahyperbolic signature),
the target space is hyperKahler.
There is a two-sphere's worth of choices of which U(1) inside
this global SU(2) is the gauged local R-symmetry.
Parametrizing this choice is the {\it twistor variable}
$\zeta\in\Pone=SU(2)/U(1)$ of self-dual geometry.
For more on the relation of twistor geometry to the 
(2,2) string, see for instance \cite{ovone}.

This self-duality structure leads to an infinite number of conservation
laws, and it is likely that this integrability underlies
the triviality of the S-matrix seen above.
Another way to exhibit these conservation laws reveals a connection
to the area-preserving diffeomorphism group $\sdiff$ \cite{husward}.
\index{area-preserving diffeomorphisms}
Consider the zero-curvature condition in two dimensions
\beq
  [\d_t+\lam A_t,\d_x+\lam A_x]=0\ ,
\label{zerocurv}
\eeq
where $\lam$ is an arbitrary (spectral) parameter, and $A_\alpha$
($\alpha=1,2$) is an $\sdiff$-valued connection.  One can represent
$\sdiff$ as the algebra of canonical transformations of a one-dimensional
phase space parametrized by coordinates $(p,q)$.  
Then \pref{zerocurv} at order $\lambda$ implies
\begin{eqnarray}
  A_t(x,t;p,q)&=&-(\d_t\d_p\Omega)\d_q + (\d_t\d_q\Omega)\d_p\nonumber\\
  A_x(x,t;p,q)&=&-(\d_x\d_p\Omega)\d_q + (\d_x\d_q\Omega)\d_p\ ;\nonumber\\
\end{eqnarray}
the $o(\lam^2)$ equations then read (in an appropriate choice
of coordinates)
\beq
  (\d_t\d_p\Omega)(\d_x\d_q\Omega)-(\d_t\d_q\Omega)(\d_x\d_p\Omega)=1\ ,
\eeq
\ie\ the Plebanski equation.  The symmetry group of \pref{zerocurv}
is the loop group of $\sdiff$; the self-dual gravity equations are
essentially the same as the two-dimensional chiral model equations
for the group $\sdiff$, where the four-dimensional spacetime is
parametrized by $(x,t,p,q)$.

\subsection{Open strings, D-branes, and matrix theory}

The connection of the target space dynamics of the (2,2)
string to $\sdiff$ leads to an intriguing toy version 
of the circle of ideas underlying matrix theory.
N=2 strings have very little dynamics; 
the N=(2,2) string has no physical transverse excitations.
The center-of-mass mode of the closed string
describes fluctuations of the \kahler\ potential 
of the self-dual metric; the corresponding mode of the open string
describes fluctuations of a self-dual Yang-Mills field.
The open string S-matrix is again essentially trivial, 
vanishing beyond the three-point function.
In such a simplified dynamical setting, one might expect to be
able to make exact statements about N=2 string D-brane dynamics as well,
perhaps even nonperturbatively.
The D-branes should strongly affect the strong-coupling behavior of the
theory, since they become light in this regime.

In a sense, in N=2 strings
self-duality plays a role similar to the BPS condition.
There is no BPS structure per se,
due to the absence of spacetime supersymmetry;
the almost topological nature of the dynamics makes up for this, however.
This is because spacetime supersymmetry of the usual superstring
is intimately related to the spectral flow that analytically relates 
the NS and R sectors in worldsheet N=2 supersymmetry \cite{specflow}.
In the N=2 string, this spectral flow is gauged, so in a sense
one has {\it only} the `BPS' sector\footnote{More precisely, the sort of
BPS states that preserve the self-dual supercharge, \ie\ $Q_\alpha$
and not $Q_{\dot\alpha}$}.  The spectral flow also generates the global
SU(2) symmetry whose preservation is tantamount to the hyper\kahler\
condition on the target space, and hence self-duality.

The fascinating proposal of matrix theory \cite{bfss}
casts the full light-cone gauge dynamics of M-theory in terms
of the quantum mechanics of the BPS sector of zero-branes.
Only the unexcited open strings stretching between the 
zero-branes are kept, yet one recaptures a remarkable amount
of the complete theory in this way.  One might hope that
N=2 string D-branes serve as a toy model of this
dynamics, since N=2 local supersymmetry permits {\it only}
such states in the physical state space of open strings --
all open string oscillations are unphysical.

Appendix B is devoted to a cursory explanation of 
the open string sector of (2,2) strings.  The upshot is
that there are D-branes, just as in the (1,1) and (0,0) strings.
\index{D-brane}
The latter lead to conventional (super)Yang-Mills
dynamics on the brane, dimensionally reduced from 10 or 26
dimensions.  Thus it is not surprising that, since the (2,2)
open string sector describes four dimensional self-dual gauge 
theory, the D-branes are governed by the dimensional
reduction of the self-duality equation:

\vskip .5cm
\begin{center}
\begin{tabular}{c c c}
${{\bf worldvolume}\atop{\bf dimension}}$ &
${{\bf integrable} \atop{\bf system}}$ &
${{\bf \scriptstyle equation}}$\\
\hline
4 & SDYM & $F+\tilde F=0$\\
3 & Bogomolny & $F_{ij} +\eps_{ijk}D_k\Phi=0$\\
2 & Hitchin & $F+[X,\bar X]=0$\\
1 & Nahm & $D_t X^i + {\eps^i}_{jk}[X^j,X^k]=0$\\
0 & ADHM & $[X^\mu,X^\nu]+{\eps^{\mu\nu}}_{\lam\sigma}[X^\lam,X^\sigma]=0$
\end{tabular}
\end{center}
\vskip .5cm

\noindent
We have denoted the branes by their total worldvolume dimensions
rather than the space dimension, since 
that notation is a bit ambiguous in signature 2+2.
Curiously, for worldvolume dimension equal to two (D-strings),
the equation is very similar to \pref{zerocurv}.
In fact one can approximate matrix theory rather closely,
in the following sense.  Consider the limit of $N$ coincident D-strings,
so that the $[X,\bar X]$ term in the Hitchin equation drops out.
then the equation governing the D-string is a zero-curvature condition
for its two-dimensional gauge field.
In the limit $N\rightarrow\infty$, one has the equations
of motion of the two dimensional chiral model with gauge group
${\su}(\infty)\sim\sdiff$; this can be shown to be equivalent
to the self-duality equations (\cf\ the last of
refs. \cite{husward}).
Thus one has a situation where one can recover the low-energy theory
corresponding to the N=2 string from the large-N limit
of one of its D-branes.  The principal difference is the lack
of a connection to the infinite momentum frame, and the corresponding
appearance of an extra dimension at strong coupling.  Also,
the transverse space to the D-branes does not appear to be related to 
the extra dimensions arising from the matrix phase space; one starts
from a two-dimensional object in a four-dimensional space whose
transverse position is nondynamical, and grows a different two dimensions
by taking the large N limit.  It may be that the appropriate starting
point is a six-dimensional theory \cite{plebsix}.
Perhaps the (0+2)-brane is close to an analogue of the D-particle
in the construction of \cite{bfss} since it spans both time directions
of 2+2 spacetime.  It would indeed be intriguing if one could 
realize self-dual gravity in terms of D-brane `constituents';
perhaps this would provide an interpretation of the `entropy'
of the nuts and bolts of compactified self-dual solutions \cite{gibhawk}.

Independent of matrix theoretic ideas, the D-branes are
the light objects of the theory at strong N=2 string coupling; they,
together with the NS instanton \pref{NSinst}, should dominate the
nonperturbative behavior of self-dual gravity.

The N=2 string D-brane system is also a convenient laboratory for the
exploration of the Nahm transform; T-duality relates the different
branes upon, for instance, compactification on $T^4$
(mirror symmetry plays an analogous role on K3).  Another intriguing
point is that the moduli space of $k$ instantons in $\su(N)$
gauge theory has been proposed as the bosonic part of the configuration
space of matrix noncritical strings \cite{matstr}.  N=2 strings generate
exactly this moduli space as the space of physical deformations,
with the D-instantons regulating the singularity at zero instanton
scale size; and manifestly incorporate T-duality.  Thus there might
be a close relationship to the dynamics of this so-far mysterious
6d noncritical string theory to quantum mechanics 
on the moduli space of N=2 strings.

%%%%%%%%%%%%%%%%%%%%%%%%%%%%%%%%%%%%%%%%%%%%%%%%%%%%%%%%%%%%%%

\section{(2,1) Strings}

\index{heterotic string}
Heterotic (2,1) strings \cite{ovtwo}
combine the self-dual, integrable structure
of (2,2) strings with the spacetime supersymmetry present in
strings with N=1 gauged worldsheet supersymmetry.  A free field 
representation for the N=1 currents is
\begin{eqnarray}
T&=&-\hf \d x\cdot\d x - \psi\cdot\d\psi -\hf \d y\cdot\d y - \lam\cdot\d\lam
\nonumber\\
G&=&\psi\cdot\d x+\lam\cdot\d y\ .
\end{eqnarray}
Here $y^a$, $a=1,...,8$ is a chiral boson and $\lam^a$ 
a Majorana-Weyl fermion.\footnote{Our conventions are as follows:
the N=1 chiral sector will be the left-moving oscillations; the
N=2 chiral sector will be right-movers.  The indices are
$\mu=0,1,2,3$ for $x^\mu\in\IR^{2,2}$; $a=1,...,8$ for $y^a\in T^8$;
often we will use a combined index $M=(\mu,a)=0,...,11$
when dealing with purely left-moving quantities.}
Using these currents as gauge constraints in conformal gauge is not
sufficient to remove both of the two timelike oscillator mode towers
of $x$ and $\psi$; one needs another bosonic and fermionic gauge current.
The simplest choice \cite{ovtwo} is a U(1) supercurrent
\begin{eqnarray}
J&=&v\cdot\d x+v_{\sst\rm int}\cdot \d y\nonumber\\
\Psi&=&v\cdot \psi+v_{\sst\rm int}\cdot \lam\ .
\label{supercurrent}
\end{eqnarray}
In order to gauge this current, it must be anomaly free, which
forces $\vv=(v,v_{\sst\rm int})$ to be a null vector: 
$v^2+v_{\sst\rm int}^2=0$.  The ghosts for $J$, $\Psi$ have $c=-3$;
therefore, they increase the critical dimension by two, so 
$d_{\rm crit}=12=10+2$.  One appears to have a `superstring in d=12',
but not really; the additional gauge constraints project
momenta and polarizations to be orthogonal to $\vv$.
If $x\in\IR^{2,2}$, then the $(y,\lam)$
system must form a $c=12$ holomorphic superconformal field theory
(just as the internal sector of the (1,0) string in $\IR^{9,1}$ must form
a $c=16$ conformal field theory).  The unique choice preserving
spacetime supersymmetry, modular invariance, etc., is for
$y$ to live on the Cartan torus of $E_8$.

Massless vertex operators are BRST equivalent to 
\begin{eqnarray}
V_{NS}^{\sst grav}&=&(\bar\Sigma_{gh}^{\sst (N=2)})
	(\Sigma_{\sst NS}^{\sst (N=1)}\,
	\xi_\mu(k)\psi^\mu) e^{ik\cdot x}\nonumber\\
V_{NS}^{\sst gauge}&=&(\bar\Sigma_{gh}^{\sst (N=2)})
	(\Sigma_{\sst NS}^{\sst (N=1)}\,
	\xi_a(k)\lam^a) e^{ik\cdot x}\nonumber\\
V_R&=&(\bar\Sigma_{gh}^{\sst (N=2)})
        (\Sigma_{\sst R}^{\sst (N=1)}\,
	u^\alpha(k) S_\alpha)e^{ik\cdot x}\ .
\label{massless}
\end{eqnarray}
As in the (2,2) case, an equivalent representative of the graviton is
\beq
  V=\int d\theta d^2\bar\theta\;e^{ik\cdot X} =
        \int d\theta d\bar\theta|_{\sst\rm N=1}\;
                I_{\mu}^\lam k_\lam\xi_\nu
			(\bar D_{\sst\rm N=1}X^\mu D X^\nu)
                        \exp[ik_\mu X^\mu+ik_a Y^a]\ ,
\eeq
so that the graviton/antisymmetric tensor has a special
polarization $h_{\mu\nu}+b_{\mu\nu}\sim I_\mu^\lam\d_\lam\xi_\nu$,
verifying the claim \pref{prepotl} of the introduction.
Similarly, the gauge field $\vareps_\mu^a\sim I_\mu^\lam\d_\lam\varphi^a$
and gravitino $\chi_\mu^\alpha\sim I_\mu^\lam\d_\lam\psi^\alpha$.
If the target space is $\IR^{2,2}\times T^8_{\rm int}$, only the states
\pref{massless} satisfy level matching 
\begin{eqnarray}
\bar L_0&=&k_\mu k^\mu =0\nonumber\\
L_0&=&k_M k^M + N_\ell=0
\label{noncompact}
\end{eqnarray}
In this case the only physical states
are at the massless level, $N_\ell=0$, much like the (2,2) string;
with no momentum in the internal $T^8$ directions, $k_a=0$.

The Virasoro/null current constraints impose restrictions
on the polarizations and momenta:

\vskip .5cm
\begin{center}
\begin{tabular}{l l l l}
Virasoro & $k_M\xi^M=0$ & $\xi\sim\xi+\alpha k$ & NS\\
	 & $\slash k u=0$ & &R\\
\medskip
Null current & $\vv_M\xi^M=0$ & $\xi\sim\xi+\alpha \vv$ & NS\\
	 & $\slash\vv u=0$ & &R\ .
\end{tabular}
\end{center}

\noindent
These constraints reduce the 12 polarizations of the NS vector
to 8 transverse, and the 32 components of the Majorana-Weyl 
R sector spinor to 8 physical states, as expected.
The spacetime interpretation of these physical states depends
on the specific orientation of the null vector $\vv$,
see figure 3.  

\vskip .5cm
{\vbox{{\epsfxsize=5in
        \nopagebreak[3]
    \centerline{\epsfbox{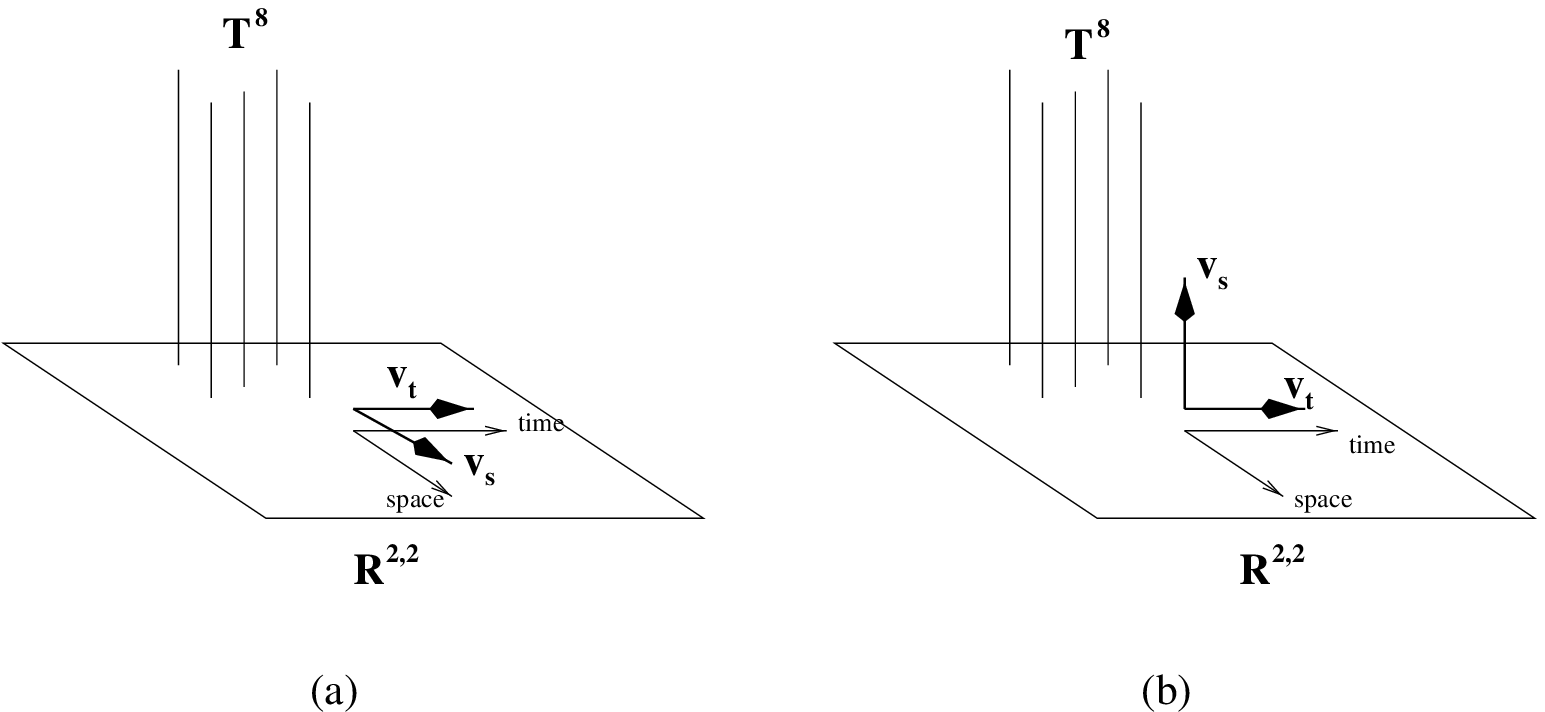}}
        \nopagebreak[3]
    {\raggedright\it \bigskip\vbox{
{\bf Figure 3.}
{\it Choices of null constraint on the left-movers.} 
 }}}}
    \bigskip}

In figure 3a, the null vector is oriented entirely within the 
$\IR^{2,2}$ base space.  The kinematics consists of 1+1 dimensional
momenta $k$ (recall that the level matching constraint eliminates
any momentum components in the $E_8$ directions), 
with the physical polarizations consisting of a
1+1 gauge field $a_\mu$, 8 scalars $\varphi^a$, and 8
fermions $\psi^\alpha$.  The spectrum is that of the 
type IIB D-string!  
\index{D-brane}
On the other hand, if the null vector
has its time component in $\IR^{2,2}$ and its space component in
$T^8$ (see figure 3b), then the kinematics is 2+1 dimensional,
and the physical polarizations are the 2+1 
gauge potential $a_\mu$, 7 scalars $\varphi^a$,
and 8 fermions $\psi^\alpha$.  The spectrum in this case is that of
the type IIA D2-brane.

One always has sixteen supersymmetries
\index{supersymmetry}
\beq
Q_\alpha=\oint\Sigma_{gh,R}^{\sst(N=1)} S_\alpha\quad,\qquad \slash\vv Q=0\ ,
\label{Qtarget}
\eeq
just as a type II D-brane breaking half the supersymmetries.
The algebra of these supercharges is (on-shell, since all considerations
are modulo BRST equivalence of the (2,1) string)
\beq
\{Q_\alpha,Q_\beta\}=(\gamma^{MN})_{\alpha\beta} P_M\vv_N\ .
\eeq
If our interpretation is correct, one would expect there to be
an additional sixteen supersymmetries which are spontaneously
broken by the brane.  These supersymmetries would be nonlinearly
realized in the worldvolume theory, and hence not visible
in the single-string Hilbert space.  One might see them in
a careful study of the vertex algebra.
In any event, the (2,1) string would appear to describe
D1- or D2-branes in static gauge, stretched across the noncompact
spatial directions of $\IR^{2,2}$.  The transverse directions
appear to be compactified on a torus, probably the $E_8$
Cartan torus.

\subsection{Toroidal compactification of (2,1) strings}

\index{compactification}
Qualitatively new features arise upon further compactification
of the (2,1) string, since the level-matching constraints
\pref{noncompact} prove to be much less restrictive.
In fact, the physical spectrum consists of a stringy
tower of `Dabholkar-Harvey' states -- ground states
on the right-moving N=2 side, oscillator excitation at arbitrary
level on the left-moving N=1 side, with momentum and winding
to compensate.  Note that since it is the worldsheet chirality
that carries spacetime supersymmetry that must be excited,
states with $N_\ell\ne0$ break target space supersymmetry;
\ie\ these states are not BPS-saturated.

Consider then the further compactification of the spatial $x^2,x^3$
coordinates of $\IR^{2,2}$, so that the target is
$\IR^2({\rm time})\times T^2\times T^8_{\rm int}$.  More generally,
one could consider an arbitrary spatial torus corresponding
to a point in the Narain moduli space $\NN^{10,2}$, but a
product torus will suffice for illustrative purposes.
The (2,1) string will in general have both momentum and winding
\beq
p^i_{\ell,r}=\frac{n^i}{R_i}\pm\frac{m^iR_i}{2}\quad,\qquad i=2,3\ .
\eeq
Reexamining the level-matching constraints
\begin{eqnarray}
0&=&-p_0^2-p_1^2+p_{2,r}^2+p_{3,r}^2\nonumber\\
0&=&-p_0^2-p_1^2+p_{2,\ell}^2+p_{3,\ell}^2+{\vec p}_{T^8}^{\;2}+2N_\ell\ ,
\label{levelmatch}
\end{eqnarray}
one sees that many more states are now available -- one need only
satisfy the mass shell condition, the level-matching constraint,
and the null constraint (as well as the highest weight condition
under the current algebra):
\begin{eqnarray}
0&=&-p_0^2-p_1^2+\biggl(\frac{n^i}{R_i}\biggr)^2+\bigl(\hf m^iR_i\bigr)^2
	+\hf{\vec p}_{T^8}^{\;2}+N_\ell\nonumber\\
0&=&2m^in^i+{\vec p}_{T^8}^{\;2}+2N_\ell\nonumber\\
0&=&\vv\cdot p_\ell\ .
\label{compact}
\end{eqnarray}
Note that the constraints now allow the internal
momenta ${\vec p}_{T^8}$ to be arbitrarily excited,
so that the kinematics is more or less ten dimensional.
These states are similar to the perturbative BPS states of the 
type II superstring; for instance, the level density
grows exponentially.  Even with this change of circumstances,
one can check that the four-point function continues to
vanish, indicating that the triviality of the S-matrix continues
to hold.  T-duality of the (2,1) string seems to imply
that there is a minimum `size' to the target circle(s) over
which the target D-brane is stretched: $R_{\rm min}\sim\lstr^{\sst (2,1)}$.

\subsection{Heterotic/type I construction}

\index{heterotic string}
It is now well-understood that the heterotic and type IA
string theories are particular asymptotic limits in the 
moduli space of $\Z_2$ orientifolds of M-theory, see figure 4.

\vskip .5cm
{\vbox{{\epsfxsize=3.0in
        \nopagebreak[3]
    \centerline{\epsfbox{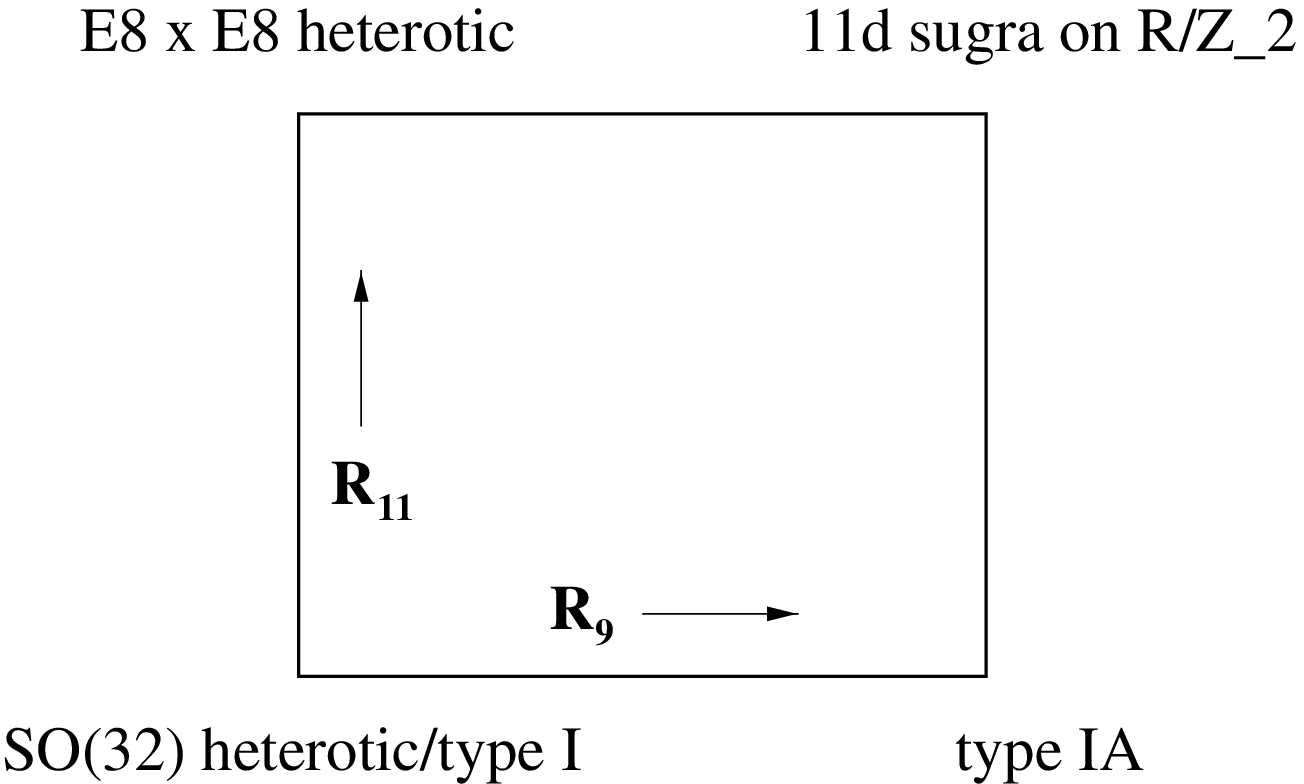}}
        \nopagebreak[3]
    {\raggedright\it \bigskip\vbox{
{\bf Figure 4.}
{\it Moduli space of $S^1\times S^1/\Z_2$ compactifications of M-theory.}
 }}}}
    \bigskip}

The simplest situation has 16 ninebranes at each of the two
orientifold planes to cancel anomalies; the low-energy dynamics consists
of 11d supergravity in the bulk, and 10d SYM on these walls.
Since the (2,1) string seems only to describe a brane in spacetime,
to realize these vacua as a (2,1) string background
one might look for a wrapped membrane/string that sees this structure,
see figure 5.  

\vskip .5cm
{\vbox{{\epsfxsize=2.5in
        \nopagebreak[3]
    \centerline{\epsfbox{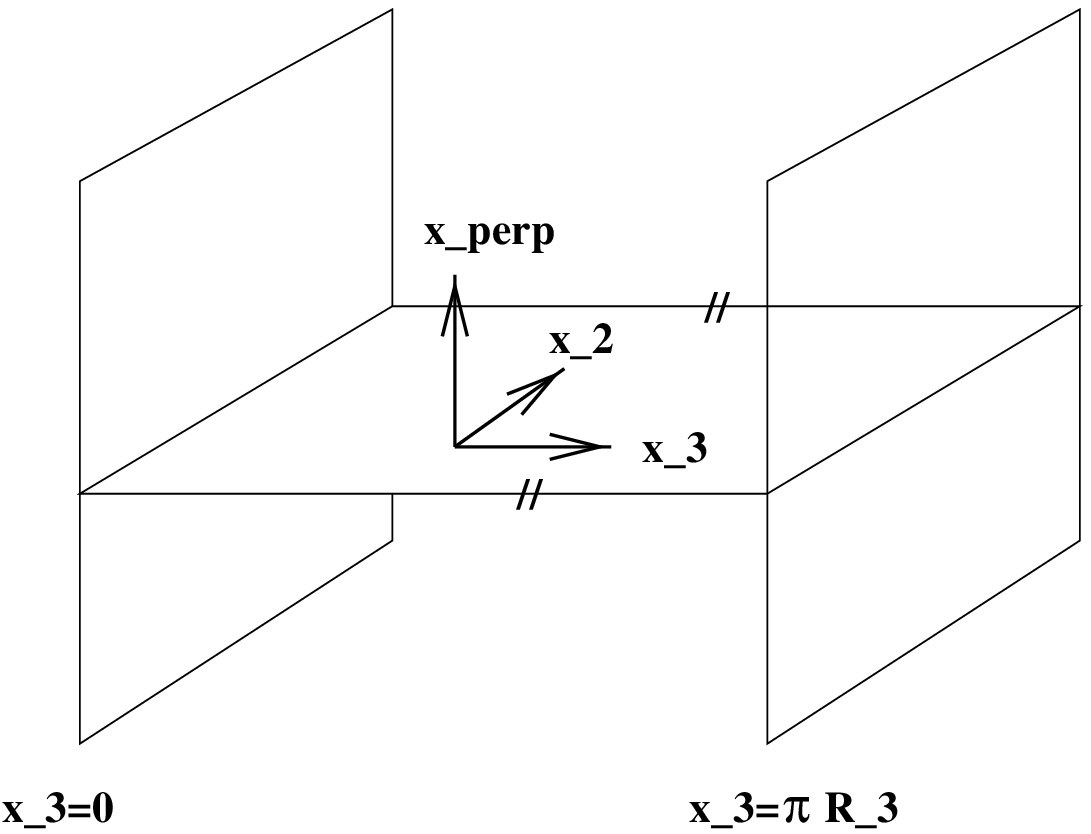}}
        \nopagebreak[3]
    {\raggedright\it \bigskip\vbox{
{\bf Figure 5.}
{\it The heterotic membrane geometry generated by the (2,1) string.}
 }}}}
    \bigskip}

In other words, one looks to describe a cylindrical, open membrane
stretched between the orientifold planes
$x_3=0,\pi R_3$.  Its boundaries
should have 1+1 dimensional fields describing rank eight current algebra,
such that one obtains a heterotic string when $R_2\gg R_3$;
on the other hand, when $R_2\ll R_3$, these fields will be frozen,
leaving the finite number of Chan-Paton labels of type IA theory.
The orientifold symmetry acts as $x_3\rightarrow -x_3$,
$A_{MNP}\rightarrow -A_{MNP}$.  The sign flip of the three-form gauge
field indicates orientation reversal of the M-theory membrane worldvolume.
Since one appears to be in static gauge, the orbifold $x_3\rightarrow -x_3$
of the (2,1) string will accomplish both the orbifold of
spacetime and the orientation reversal of the brane worldvolume.
One simply wants to find a consistent supersymmetric orbifold
of the (2,1) string with this operation as part of the orbifold
group.  A straightforward chain of logic leads to 
a unique answer satisfying the orbifold level-matching constraints, 
preserving half the sixteen spacetime supersymmetries, 
and treating all the internal coordinates $y$ on the same footing.
Right-moving worldsheet supersymmetry is preserved only if,
in addition to $x_3\rightarrow -x_3$, one simultaneously
flips $\psibar_3\rightarrow -\psibar_3$.  Now consider the right-moving
U(1) R-current, \eg\ $\bar J=\psibar^0\psibar^3+\psibar^1\psibar^2$;
there are two choices for the $\Z_2$ to have a well-defined action
on the gauge algebra: (1) $\psibar_0\rightarrow-\psibar_0$, which one
can show does not lead to a spacetime supersymmetric solution;
and (2) $\psibar_1\rightarrow-\psibar_1$, which gives the 
twisted N=2 algebra \cite{twistedNtwo}
\beq
\bar J\rightarrow -\bar J\quad,\qquad\bar G^+\leftrightarrow \bar G^-\ .
\eeq
Proceeding along these lines, one finds a unique $\Z_2$
satisfying the above requirements:

\begin{center}
\begin{tabular}{l@{$\rightarrow\ $}l}
$(x_1,x_3;y_1,..,y_8)$ & $-(x_1,x_3;y_1,..,y_8)$ \\
$(\psi_1,\psi_3;\lam_1,...,\lam_8)$ & $-(\psi_1,\psi_3;\lam_1,...,\lam_8)$ \\
$(\psibar_1,\psibar_3)$ & $-(\psibar_1,\psibar_3)$ \ ,
\end{tabular}
\end{center}
with all other coordinates invariant.

Vertex operators describe physical fluctuations of the target brane,
and have the form $\OO \cos[k_3x^3]$ if the polarization operator $\OO$
is $\Z_2$ even, and $\OO\sin[k_3x^3]$ if $\OO$ is $\Z_2$ odd.
At the massless level, one finds Neumann boundary conditions 
($\cos[k_3x^3]$) for polarizations along $(0,2;4,...,11)$, and
Dirichlet boundary conditions ($\sin[k_3x^3]$) 
for polarizations along $(1,3)$.  Thus, in space there are nine Neumann
and one Dirichlet boundary condition; in time, one Dirichlet 
and one Neumann.  Since we
wish to avoid thinking about what a Dirichlet boundary condition
means in physical time, we place the time component of the vector
$\vv$ (defining the null projection $J$) 
in the $x_1$ direction; thus its space component
must also be one of the $\Z_2$ twisted directions $(2,4,...,11)$.
Choosing the latter in one of the internal directions,
one finds now an open membrane stretched between eight-branes
in a type IIA description; that is, when the target $(x_2,x_3)$
torus orbifold is a large cylinder, in the low-energy limit
one has a 2+1 target space gauge field
coupled to 7 scalars with Neumann boundary conditions.
This is precisely the bulk dynamics of an open membrane stretched
between D8-branes.  One can lift to this effective
theory to eleven dimensions 
by dualizing the vector to another scalar (which one easily
checks also has Neumann boundary conditions) to find an open
membrane stretched between orientifold ninebranes.

The boundary dynamics comes from the twisted sectors of the
(2,1) string orbifold, as these states are pinned to the
fixed points $x_3=0,\pi R_3$ which are the orientifold planes.
Since the orbifold is an asymmetric one, the number of
states at each of the two fixed points $x_3=0,\pi R_3$
is the square root of the number of fixed points of the
internal $T^8/\Z_2$ orbifold, $\sqrt{2^8}=16$.
Consistency of the operator algebra shows that these states appear 
in the Ramond sector, and an analysis of the BRST constraints
shows that they are chiral in the target space.  Thus
there are sixteen massless fermion fields living on each boundary
of the open target membrane, which describe a rank eight
current algebra, and are inert under spacetime
supersymmetry.  
Half the supersymmetry charges \pref{Qtarget} 
are broken by the orbifold.
Note that this is exactly the spectrum
found on the D8-brane boundaries of the matrix theory description
of the heterotic string, a fact that will be important for us below.

There is a very similar orbifold \cite{kmtwo}, also breaking
half the supersymmetry, obtained if one relaxes the condition that all
internal coordinates are treated identically. 
Twisting half rather than all the $y^a$, $\lam^a$ 
\begin{center}
\begin{tabular}{l@{$\rightarrow\ $}l}
$(x_1,x_3;y_1,..,y_4)$ & $-(x_1,x_3;y_1,..,y_4)$ \\
$(\psi_1,\psi_3;\lam_1,...,\lam_4)$ & $-(\psi_1,\psi_3;\lam_1,...,\lam_4)$ \\
$(\psibar_1,\psibar_3)$ & $-(\psibar_1,\psibar_3)$ \ ,
\end{tabular}
\end{center}
is also a consistent asymmetric orbifold.  The twisted sector
ground states transform as a hypermultiplet under the
eight remaining supersymmetries. 

Finally, it should be noted that we have only described the
low-energy spectrum; there is again a stringy tower of left-moving
states just as in the untwisted, toroidally compactified (2,1) string.

\section{S-matrix, effective action and geometry of (2,1) strings}

\index{effective action}
The tree-level, three-point S-matrix factorizes between 
left- and right-movers, hence we can immediately deduce the 
answer by combining (2,2) string result \pref{Stwotwo} with that
of the superstring; for three massless metric perturbations,
the result is
\beq
\vev{V_\xi(1)V_\xi(2)V_\xi(3)}=
	(k_1\cdot I\cdot k_3)_r(\xi_1\cdot\xi_2\;k_1\cdot\xi_3)_\ell
		+{\rm cyclic}\ ,
\label{bosthreept}
\eeq
leading to the cubic effective Lagrangian
\beq
\LL_{\rm eff}^{\sst(3)}=\hf F_{\mu\nu}F_{\nu\lam}F_{\lam\sigma}F_{\sigma\mu}
	-\coeff18(F_{\alpha\beta}F_{\beta\alpha})(F_{\mu\nu}F_{\nu\mu})
\label{Lthree}
\eeq
with $F_{\mu\nu}=\d_{(\mu}\xi_{\nu)}$.  Curiously, this is a term
in the expansion of the Born-Infeld Lagrangian around
the background $F_{\mu\nu}=I_{\mu\nu}$.  One can incorporate
the effects of the internal scalars $\varphi$ simply by letting
the polarization indices in \pref{Lthree} run over 10d (or 12d, since we
are imposing the null constraints by hand), while keeping
the kinematics 1+1 or 2+1 dimensional.
Regarding the fermions, their S-matrix is
\beq
\vev{V_\psi(1)V_\xi(2)V_\psi(3)+V_\psi(1)V_\varphi(2)V_\psi(3)}=
	(k_1\cdot I\cdot k_3)_r
	[\bar u_1(\xi^\mu\Gamma_\mu+\zeta^a\Gamma_a)u_3]_\ell\ ,
\eeq
where $\Gamma_M$ are twelve dimensional gamma matrices.
One obtains the effective Lagrangian
\beq
\LL_{\rm eff}^{\sst(3)}=
	(\psibar\Gamma^\mu\d_\nu\psi)I^{\nu\lam}F_{\lam\mu}
	+(\psibar\Gamma^a\d_\nu\psi)I^{\nu\lam}\d_\lam\varphi^a\ .
\label{Lthreeferm}
\eeq
These are the only nonzero S-matrices%
\footnote{Hence the analogy with the 1+1d noncritical
string made in the introduction, wherein the `bulk' S-matrix is
also trivial.}
in $\IR^{2,2}\times T^8$,
but this does not mean that we have found the full effective action.
It happens that the iteration of these cubic field-theoretic
vertices generates a four-point S-matrix, such that kinematics
in $\IR^{2,2}$ allows the pole terms to cancel among $s$, $t$, and
$u$ channels, leaving an explicit four-point contact term.
For the full S-matrix to vanish, one must add a cancelling
quartic term in the effective action.  Continued iteration
yields terms to all orders in small fluctuations.  This 
situation contrasts with the (2,2) string, 
where the symmetry is powerful enough to
cause the potential four-point contact term to vanish, 
and allows the effective action to cubic order to 
in fact be the exact theory.

The simplest way to extract the full answer is to realize that
the iteration of the cubic terms will generate contact terms
of the form $k^n(\delta\phi)^n$ at $n^{\rm th}$ order in fluctuations,
while fluctuations in the geometrical fields $h,b,\vareps$
are all of the form $\d(\delta\phi)$ due to the (2,1) supersymmetry
constraints.  Hence the effective action found from
the beta function will come entirely from one loop (higher
loops will come with extra factors of $\alpha'k^2$, and hence 
more powers of momenta than small fluctuations).
The beta function equations \pref{betafneq} can be integrated
once to give
\beq
\Gamma_\mu\equiv \Gamma^{\rho}_{\nu\mu}I^\nu_\rho=0
\eeq
with the effective action
\beq
S_{\rm eff}^{\rm grav}=
	\frac1{g_{\rm str}^2}\int d^4x\;\det^{\half}[\eta_{\ij}+F_\ij]\ ,
\eeq
at least for the gravity sector.  The gauge fields may be
added via an analysis of sigma model anomalies \cite{kmthree},\cite{hull};
the result is 
\beq
S_{\rm eff}=
	\frac1{g_{\rm str}^2}\int d^4x\;\det^{\half}
		[\eta_{\ij}+F_\ij+\alpha'\d_i\varphi^a\d_\jbar\varphi^a]\ .
\label{dbi}
\eeq
It has been checked \cite{kmthree}
that this action reproduces all cubic
and induced quartic vertices involving only bosons,
and that the S-matrix vanishes at quartic order.
The action is remarkably similar to the static gauge effective
action for D-branes.  
\index{D-brane}
There are two significant differences, however:
(1) For target space $\IR^{2,2}\times T^8$, it is exact
to all orders in $\alpha'=\lstr^2$, presumably related to the fact
that there are no oscillator excitations of the underlying (2,1) string;
and (2) the dynamics is integrable.  The vanishing of the
S-matrix in the fully compactified case indicates that 
this integrability continues to hold  even when the kinematics
is essentially ten-dimensional.  It would be very interesting
to work out the generalization of \pref{dbi} in this case.
It would also be interesting to explore the possibility of nontrivial
solutions along the lines of \pref{NSinst}.
 
The terms in the effective action involving fermions should
in principle be determined by supersymmetry.  However,
it is difficult to compute quantities involving nontrivial Ramond 
vertex backgrounds (there is no sigma-model approach).
A better understanding of the spacetime supersymmetry
current algebra and its anomaly structure would be helpful,
since it was essentially this structure which enabled us to determine
the dependence of the effective action on the scalars $\varphi^a$.
As mentioned in the introduction, the physical fields on
the target brane are all Nambu-Goldstone modes of the 
spacetime symmetries spontaneously broken by the brane,
and therefore the effective action is expected to be largely 
determined by the various broken and unbroken symmetries.
Because the bosonic structure is so similar to the 
Dirac-Born-Infeld/Nambu-Goto action, it is natural to guess
that the full answer including fermions has a structure similar
to that of D$p$-branes in static gauge
\begin{eqnarray}
S^{\sst(p)}&=&\int d^{p+1}\sigma\;\det^\half\biggl[\eta_{\mu\nu}+F_{\mu\nu}
	+\d_\mu\phi^a\d_\nu\phi^a\label{staticbrane} \\
	& &\hskip 3cm
	-2\psibar(\Gamma_\mu+\Gamma^a\d_\mu\phi^a)\d_\nu\phi
	+(\psibar\Gamma^a\d_\mu\psi)(\psibar\Gamma^a\d_\nu\psi)\biggr]\ .
	\nonumber
\end{eqnarray}
The sorts of terms appearing in the
expansion of this action are compatible with
with the cubic S-matrix \pref{Lthreeferm}, but the full
structure is far from understood.  This fermionic completion
of of the action \pref{dbi} represents a rather nontrivial
coupling of SDYM to self-dual gravity with torsion, with an
additional fermionic symmetry whose geometry ought to
be quite intriguing.

\section{Connections between (2,1) strings and matrix theory}

\index{matrix theory}
In this final section of these lectures, I would like to present
some evidence for the idea that (2,1) strings are closely
related to the formulation of M-theory on $T^9$ 
as a matrix model.

A striking feature of supergravity U-duality is the appearance of
exceptional groups in less than six noncompact spacetime dimensions.
Naive extrapolation leads to a duality group $E_{9(9)}(\Z)$
in two dimensions \cite{hulltown}.  
However, the solutions to low-energy supergravity
which exhibit the various BPS charges permuted under duality,
become more and more singular the lower the dimension.
For instance, in 2+1 dimensions, $n$ fundamental strings wound around
a given cycle of the internal $T^7$ has a dilaton background
of the form \cite{dabharv}
\beq
e^{-\Phi}=e^{-\Phi_0}-
	\sum_{i=1}^n 8G_N\lstr^{-2} {\it ln}|\vec r-{\vec r}_i|\ ;
\eeq
one can introduce an arbitrary number of such sources, but
the dynamics becomes strongly coupled near the core of each one.
Presumably the correct solution receives modifications due to
the nonperturbative states that become light there 
(\cf\ \cite{ghm}).
This is supported by the observation that, in the IIB theory,
one can wrap a D7-brane around the $T^7$; 
these are in the same U-duality multiplet
as the fundamental string, yet they have a rather different
classical solution -- the core is nonsingular (from the
viewpoint of F-theory), but one cannot introduce more than 24 sources.
In 1+1 dimensions, the situation is even worse \cite{banksuss};
any source generates a response in the dilaton of the form
\beq
e^{-\Phi(x^+)}=e^{-\Phi_0}-\int\int^{x^+}_{-\infty} T_{++}\ ;
\eeq
positive stress-energy of the source forces a singularity
in the dilaton at finite $x^+$, so again information of a more
nonperturbative nature is needed.
Finally, if one considers the velocity-dependent forces between
these objects in low dimensions, eventually they become confining
even between the BPS states.  For instance, in matrix theory
the velocity dependent force between two matrix partons on 
$\IR^{10-d,1}\times T^d$ is $v^4/r^{7-d}$, becoming confining
below 3+1 dimensions.  

It is not clear what to make of these observations.  
Two possibilities are
\begin{itemize}
\item
Supergravity makes sense as a low-energy theory,
but that the branes that have been our guide to defining
matrix theory do not make sense -- or at least cannot be present
in the arbitrary numbers needed to define a large-N continuum 
limit.\footnote{Also the branes have a strong back-reaction on
spacetime, perhaps precluding the existence of the null Killing
vector needed to define the light cone gauge.}
\item
Matrix theory exists in all dimensions; however, the confining
potential between matrix partons means that there
will be no moduli space for the gauge dynamics
defining matrix theory in low dimensions, therefore no
low-energy approximation that one could call spacetime,
therefore no supergravity.
\end{itemize}

\noindent
The connection between (2,1) strings and brane dynamics cries
out for some connection to M-theory.
Could (2,1) strings describe the `low-dimensional phase'
of matrix M-theory conjectured above?  
There are a number of reasons to think so:
\begin{enumerate}
\item
At least eight coordinates are compact on the scale $\lstr^{\sst(2,1)}$.
\item
Massless physical states of the fully compactified
(2,1) string have interactions similar to 
what one would expect from SYM with gauge group $\sdiff$.\footnote{In
matrix theory on $T^9$, the theory cannot quite be SYM even
at low energies, due to anomalies.}
\item
The $\Z_2$ orbifold that gives the heterotic string acts as expected
in matrix theory \cite{hetrefs}.
\item
In the fully compactified (2,1) string, the decompactification limits
are analogous to the limits which yield matrix IIA/B strings.
\end{enumerate}

Consider again the vertex operators of the fully compactified (2,1) string
\begin{eqnarray}
V_{NS} &=& ({\rm ghosts})\bigl(\xi_M(p_\ell)\psi^M\bigr)\, 
	\exp[ip_\ell\cdot x_\ell+ip_r\cdot x_r] 
		\equiv \AA_M(x_\ell,x_r) \nonumber\\
V_R&=& ({\rm ghosts}) \bigl(u_A(p_\ell)S^A\bigr)\,
	\exp[ip_\ell\cdot x_\ell+ip_r\cdot x_r] \equiv \lam_A(x_\ell,x_r) \ .
\end{eqnarray}
The three-point function of, for instance, the bosons \pref{bosthreept}
generates the cubic coupling
\beq
\LL^{\sst(3)}=g_{\rm str}\;\d_\ell^M\AA^N\{\AA_M,\AA_N\}\ ,
\label{lthreemat}
\eeq
where $\{F,G\}=(\d F/\d x_r^\mu)I^{\mu\nu}(\d G/\d x_r^\nu)$
is a Poisson bracket induced by the complex structure of the
right-moving N=2 supersymmetry.  This cubic vertex has the
same structure as Yang-Mills with a gauge group of symplectic
diffeomorphisms.  In this interpretation, we wish to regard 
$x_\ell$ as transverse `spacetime' coordinates, and $x_r$ as
phase space coordinates on the Lie algebra of $\sdiff$
\`a la matrix theory.  In fact, the left- and right-moving degrees
of freedom are not independent; they are coupled through the level-matching
constraints \pref{levelmatch}.  Moreover, the lattice of
momenta $\Gamma^{10,2}$ is generically not factorized
as $\Gamma^{10}_\ell\times\Gamma^2_r$.  Nevertheless,
there may be a sense in which a structure of $\sdiff$ gauge
theory is present off-shell \cite{givshap}.  
It is possible that this gluing of gauge and spacetime
fluctuations is an artifact of the expansion about a 
particular classical solution, much as a monopole ties rotations
in space and isospin.

\index{area-preserving diffeomorphisms}
To see the action of $\sdiff$,
consider the right-moving chiral part of the vertex operators, 
taking sets which are mutually
local in their OPE's; this means
\beq
p_i\ ,\quad i=1,2,3\ :\qquad p_i^2=\sum_{i=1}^3 p_i=0\ .
\eeq
The generic construction of such sets is as follows: Take two
orthogonal null vectors $n_{\sst(1)},n_{\sst(2)}$ (\ie\
$n_{\sst(1)}^2=n_{\sst(2)}^2=n_{\sst(1)}\cdot n_{\sst(2)}=0$,
and $n_{\sst(1)}\ne\alpha n_{\sst(2)}$).  This is possible in 
signature 2+2.  The $n_{\sst(a)}$ span a {\it self-dual null plane} $\NN$.
Now focus on momenta $p=a n_{\sst(1)}+b n_{\sst(2)}$ 
which are arbitrary linear combinations of the two basis vectors.
The chiral vertex operators $J_p=\oint dz\int d^2\thetabar 
\;\exp[ip\cdot x_r]$
satisfy the algebra
\beq
[J_p,J_q]=p\cdot I\cdot q \; J_{p+q}
\label{sdiffalg}
\eeq
modulo BRST equivalence.  
Momenta $p\not\in\NN$ do not have mutually local OPE's,
and hence no well-defined algebra structure.
These on-shell $p_r\in\NN$ are completely
determined by their spatial components; the null condition forces
the length of the spacelike part to equal that of the timelike part,
so that the triple $(p,q,-p-q)$ form congruent triangles in the temporal
$x_0$-$x_1$ and spatial $x_2$-$x_3$ planes 
(an overall relative orientation of these
two triangles is determined by the choice%
\footnote{This choice
is not unique; in fact the set of self-dual null planes is
parametrized by a coordinate $\zeta\in\IR{\bf P}^2$ which is 
the twistor parameter.} of $\NN$).
The upshot is that the algebra \pref{sdiffalg} is an algebra of
two-dimensional and not four-dimensional symplectic diffeomorphisms,
\ie\ $\sdiff\sim\su(\infty)$.

Thus the structure of the fully compactified (2,1) string strongly
resembles that of matrix theory on $T^9$, although there are
profound differences due to the constraints relating left- and
right-moving coordinates $(x_\ell,x_r)$.  Further support for
this idea is provided by the fact that the decompactification
limit of the (2,1) string is described by a D-brane style Lagrangian
for either strings or membranes (depending on the orientation
of the null vector $\vv$); in fact it is precisely the D-string
action if the null vector $\vv\in\IR^{2,2}$.%
\footnote{The left-right
constraints are most powerful here, since they restrict all
momenta to $p_\ell=p_r$.  It would seem then that this limit 
describes only a single D-string and not $N\rightarrow\infty$ of them.
It may be that this freezing of U(N) down to U(1)
is an artifact of the string perturbation expansion, which
takes place about a particular classical solution;
perhaps this symmetry is broken in the particular vacuum
seen by the (2,1) string.}
D-string dynamics is precisely what one finds in matrix theory
if one shrinks a compactified circle to a radius $R_i\ll\lpl$;
then the dual circle on which the SYM dynamics takes place 
lives on a circle of radius $\Sigma\sim\lpl^3/R_{i}R$,
so that the IR dynamics becomes 1+1d SYM in the limit \cite{matstr}.

The heterotic/type I construction in section 3.2 follows precisely
the pattern one would expect of matrix theory on $T^9$.
In matrix theory, the heterotic/type I theory on $T^d$ arises 
from a SYM orientifold on ${\tilde T}^d\times{\tilde S}^1$;
the orientifold $\Z_2$ acts simultaneously on the torus
and acts as the involution that sends $SU(N)\rightarrow SO(N)$.
One must also add by hand 32 fermionic matter multiplets
in order to cancel gauge anomalies; these represent the couplings
of the zero-brane/supergravitons to the nine-branes
of the background spacetime.  All of this structure has a direct
parallel in the (2,1) string.  The orbifold group of section 3.2
acts on the left-moving `spacetime' coordinates $x_\ell$
by reflecting nine spatial coordinates (of which one is removed
by null projection) while leaving one untouched --
in other words, as though the parameter space were 
${\tilde T}^8/\Z_2\times{\tilde S}^1$.  Simultaneously, it acts
to flip the right-moving U(1) R-current $\bar J\rightarrow -\bar J$.
Thus the $\sdiff$ algebra \pref{sdiffalg} will undergo a $\Z_2$
quotient which one can perhaps think of as $SO(\infty)$.
Finally, one finds the requisite 32 fermionic states in the
twisted sector (and here they are not simply put in by hand).
Decompactification of a circle again resembles the construction
of the matrix heterotic string \cite{hetrefs}, although again
the left-right gluing hides the gauge group structure in this limit.

The alternate orbifold presented at
the end of section 3.2 seems to be 
a ${\tilde T}^4/\Z_2\times{\tilde T}^5$ orbifold,
but rather than sixteen hypermultiplets appearing in the 
twisted sector (as expected in matrix theory on
this space), we find only one.

At this point several remarks are in order.  First, the (2,1) string
appears to lack the parameters needed to explore
the full 128-parameter moduli space of matrix theory on $T^9$.
Eight circles are fixed to the scale $\lstr^{\sst(2,1)}$, so
it is not clear exactly how to connect the construction to the
rest of M-theory.  This aim might be furthered by understanding
how to construct U-duality multiplets in the context of (2,1)
string theory.  The perturbative string states will be the momentum
states of the SYM theory under the present interpretation
(which are longitudinally wrapped membranes in matrix theory).
The duality group is probably $E_{9(9)}(\Z)$ \cite{hulltown}
and does not act entirely within the perturbative theory.%
\footnote{This corrects an erroneous interpretation in \cite{martinec},
where an attempt was made to connect the duality group
to the symmetries of the internal $E_8$ torus.}
In fact, one expects \cite{egkr} states whose masses scale as
arbitrarily high powers of $1/\gym^2$.
Since $\gym\propto g_{\rm str}$, we are not going to find 
these easily.

%%%%%%%%%%%%%%%%%%%%%%%%%%%%%%%%%%%%%%%%%%%%%%%%%%%%%%%%%%%%%%
\vskip 1cm
\begin{center}
{\bf Appendices}
\end{center}
\vskip .5cm
%%%%%%%%%%%%%%%%%%%%%%%%%%%%%%%%%%%%%%%%%%%%%%%%%%%%%%%%%%%%%%

\appendix
\section{Technology of local N=2 worldsheet supersymmetry}

The ghost operators in superstring vertices provide the
appropriate superconformally covariant geometrical measure 
for their integration over the worldsheet
(for details, see \cite{fms,gidmart}).
The superghosts have a number of different `pictures', or
BRST representatives, because the splitting of the supersymmetry
ghosts into creation and annihilation operators
is ambiguous.  For the N=1 superconformal algebra, there is a single
pair of such ghosts $\beta$, $\gamma$; for N=2 one has 
two -- $\beta_\pm$, $\gamma_\pm$ for the two supersymmetries $G_\pm$.
The ghost current may be bosonized as\footnote{For the twisted N=2
algebra, it is more convenient to bosonize the twist eigenstates
$\beta_+\gamma_+\pm\beta_-\gamma_-$ rather than the charge eigenstates.}
$\d\phi_\pm=\beta_\pm\gamma_\pm$.  The exponentials
$exp[\alpha\phi_\pm]$ interpolate between the various vacua;
$\phi_\pm$ are free fields.

In the NS sector there are two canonical pictures for a given chiral
sector of the usual N=1 superstring; the N=2 string doubles this.
Consider a superfield, with expansion in components
\beq
\OO=\OO^{\sst(0)}+\thetabar_+\OO^{\sst(1)}_-
	+\thetabar_-\OO^{\sst(1)}_+ + \thetabar_+\thetabar_-\OO^{\sst(2)}
\eeq
under the right-moving N=2 supersymmetry.  For example, the exponential
$exp[ik\cdot X]$ expands as
\begin{eqnarray}
e^{ik\cdot X}&=&e^{ik\cdot x}\bigl[
	1 + i\thetabar_+\; k\cdot(\One+I)\cdot\psibar
	 + i\thetabar_-\; k\cdot(\One-I)\cdot\psibar\\
	& & + \thetabar_+\thetabar_-\bigl(k\cdot I\cdot\dbar x-
		(k\cdot\psibar)(k\cdot I\cdot\psibar)\bigr)\bigr]\ .
\end{eqnarray}
BRST invariant vertex operators are then
\begin{eqnarray}
V^{\sst(0)}&=&e^{-\phibar_+-\phibar_-}e^{ik\cdot x}\nonumber\\
V^{\sst(1,\pm)}&=&e^{-\phibar_\pm}[k\cdot(\One\mp I)\cdot \psibar]
	e^{ik\cdot x}\nonumber\\
V^{\sst(2)}&=&\bigl[k\cdot I\cdot\dbar x-
                (k\cdot\psibar)(k\cdot I\cdot\psibar)\bigr]
		e^{ik\cdot x}\ .
\end{eqnarray}
Any combination of these with ($\pm$) ghost charges adding
up to $-2$ (to cancel the background charge on the sphere)
will yield the same on-shell S-matrix amplitude \pref{Stwotwo}.
The N=1 structure is as usual \cite{fms} in terms
of the bosonization $\d\phi=\beta\gamma$, with additional
ghost factors in the Ramond sector from the spin field for
the supercurrent $\Psi$, equation \pref{supercurrent}.
These are again constructed in the standard way: the corresponding spin 1/2
ghosts $\tilde\beta$, $\tilde\gamma$ are bosonized
via $\d\rho=\tilde\beta\tilde\gamma$, and the standard picture for
the Ramond sector vertices involves a factor
$\Sigma_{gh,R}^{\sst (N=1)}= e^{-\half\phi+\half\rho}$.
The NS vertex operators in equations \pref{Vtwotwo},\pref{massless} 
have been written in the zero picture for both left- and right-movers.

%%%%%%%%%%%%%%%%%%%%%%%%%%%%%%%%%%%%%%%%%%%%%%%%%%%%%%%%%%%%%%

\section{(2,2) open strings and D-branes}

%%%%%%%%%%%%%%%%%%%%%%%%%%%%%%%%%%%%%%%%%%%%%%%%%%%%%%%%%%%%%%
The open string sector describes self-dual Yang-Mills, in a formulation
due to Yang \cite{ovone},\cite{marcusopen}.  The equations of motion are
\begin{eqnarray}
  F^{(2,0)}&=&F^{(0,2)}=0\nonumber\\
  k\wedge F&=&0
\label{sdym}
\end{eqnarray}
The single physical string mode is again the center-of-mass, which is
an adjoint scalar $\varphi$ related to the gauge field by
\beq
  A=(D e^{-\varphi}) e^\varphi\quad,\qquad 
  \bar A=(\bar D e^\varphi)e^{-\varphi}\ ,
\label{sdgauge}
\eeq
with $D$ a background covariant derivative.
This ansatz automatically solves the (2,0) and (0,2) components of
\pref{sdym}, and converts the (1,1) part 
into a wave equation for $\varphi$.
In the modern interpretation of gauge charge, the 
Chan-Paton indices are carried by branes
filling spacetime, whose number may be
determined by closed string tadpole cancellation (of course, such 
restrictions are not active at open string tree level, 
where we can formally consider any classical Chan-Paton group).
A calculation of the tadpole \cite{marcusopen} indicates a gauge group
G=SO(2); however, it should be noted that there are many massless tadpoles
in N=2 string loop amplitudes, whose interpretation is currently
unclear, rendering the determination of $G$ somewhat ambiguous.

\subsection{Boundary conditions and D-branes}

\index{D-brane}
The open string boundary conditions of N=2 superconformal field theory
have been investigated by \cite{ooguri}.
They fall into two classes:
\begin{eqnarray}
\label{BCs}
	{\rm A-type:}\qquad& & J_\ell=-J_r\ ,\quad 
		G_\ell^+=\pm G_r^- \nonumber\\
	{\rm B-type:}\qquad& & J_\ell=+J_r\ ,\quad G_\ell^+=\pm G_r^+ \ , 
\end{eqnarray}
where $d$ is the complex dimension of the target space and $p+1$
is the brane dimension\footnote{Unfortunately, 
this convention is not particularly
well-adapted to N=2 strings with signature 2+2; nevertheless we will
adhere to it, using the notation ($s$+$t$)-brane for a brane with $s$
space and $t$ time dimensions when necessary.}.
These two types of boundary condition are related to the two
types of N=2 superfield \cite{ghr}: If a chiral field obeys type B 
boundary conditions, then its T-dual is a 
twisted chiral superfield satisfying type A boundary conditions,
and vice-versa.
Also, for a single free
superfield, type B boundary conditions arise when the scalar components
are either both Neumann or both Dirichlet, whereas type A has one
Dirichlet and one Neumann boundary condition.

The dynamics of type II branes in ten dimensions is, at long wavelengths,
the dimensional reduction of the Yang-Mills theory governing 10d
open strings, the dynamics of the Dirichlet coordinates being ``frozen''
on the brane.  Correspondingly, one expects 
the dynamics of N=2 string D-branes
to be governed by the dimensional reduction of 
the self-dual equations \pref{sdym}.
For instance, the D-string equations are the Hitchin equations
\beq
  \bar D_A X=D_A\bar X=0\quad,\qquad F_A+[X,\bar X]=0\ .
\label{hitchineq}
\eeq
Here the dynamical field is a 2d scalar $\varphi$ in the
adjoint of U($N_2$) determining the reduced gauge field \pref{sdgauge}.
The covariantly constant (and therefore nondynamical)
Higgs field $X$ describes the transverse
positions of the $N_2$ D-strings.  
When $N_2$ D-strings coincide, these equations become the 2d chiral
model equations.
Similarly, one can show that
the D-particle is described by the Nahm equations
\beq
  DX^i=\eps^{ijk}[X^j,X^k]\ ,
\label{nahmeq}
\eeq
and the D2-brane by the Bogomolny equations
(for recent related work, see \cite{leeyi})
\beq
  F_{ij}=\eps_{ijk}D_k X\ .
\label{bogoeq}
\eeq
There are other possible reductions, for instance imposing Dirichlet
boundary conditions along a null direction, or wrapping branes
around homology cycles of nontrivial self-dual four-manifolds such
as K3; we will not discuss them here.
Note that the equations \pref{hitchineq}-\pref{bogoeq} are all in the family
of Uhlenbeck-Yau type equations describing 
BPS configurations of D-branes (\cf\ \cite{harvmoore}),
another indication of the resemblance of the dynamics to
the BPS sector of superstrings.

The D-instanton is also of some interest, in that its amplitudes may
serve to define an off-shell continuation of N=2 string
dynamics, and thus an off-shell quantum theory of
self-dual gravity and Yang-Mills.  
These amplitudes are in some respects similar to
those of macroscopic loops in noncritical string theory 
\cite{macroscopic}.

One may also consider composite systems of $p$- and $p'$-branes.
For example D-instantons in the open string theory are
composites of a ($0$+$0$)-brane and a ($4$+$0$)-brane
which regularize an abelian instanton
\footnote{Assuming the tadpole cancellation giving gauge group SO(2);
otherwise -- for instance at open string tree level --
it may be consistent to consider multiple ($4$+$0$)-branes
and thus nonabelian (\eg\ SU($\rm N_{4+0}$)
self-dual YM gauge group.  Then the D-instantons
carry fundamental representation gauge quantum numbers.}.
The $0$-$4$ strings should regulate the moduli space
of ADHM instantons (although this needs to be checked).

\subsection{Coupling D-branes to self-dual gravity}

The SDYM open string dynamics couples to the self-dual
gravity of N=2 closed strings, since as usual open string loops
factorize on closed string exchange.  
Moreover, performing a \kahler\ gauge transformation 
$\delta K=\Lambda(X)+\Lambda^*(X^*)$ on the worldsheet
action
\beq
  S=\int_\Sigma d^2z d^2\theta d^2\thetabar \;K(X,X^*)
	+ \oint_{\d\Sigma} ds d^2\theta \;\varphi(X,X^*) \ ,
\eeq
one sees that $\Lambda$ can be pushed onto gauge transformations
of the U(1) part of the prepotential $\varphi$
which encodes the self-dual U(N)=SU(N)$\times$U(1) open
string gauge background \pref{sdgauge}.

Marcus has shown \cite{marcusopen}
that the leading correction to the closed string equation
of motion \pref{gfixed} is
\beq
  \det[g_{\ij}]=1 - \frac{2\kappa^2}{g^2}{\rm Tr}[F_\ij F^\ij]
	+O(g^4)\ ,
\eeq
where $g$ and $\kappa$ are the open and closed string couplings,
related by $\kappa\sim\sqrt\hbar g^2$.
Using the open string equation of motion $k\wedge F=0$, one may
rewrite this in a way that manifests the above gauge symmetry:
\beq
  \coeff1N{\rm Tr}[(k+gN^{\frac12}F)\wedge(k+gN^{\frac12}F)]=
	\Omega\wedge\Omega^*\ 
\eeq
where $\Omega$ is the holomorphic two-form.

It seems, however, that D-branes will not act 
as linearized sources for static gravitational
and/or axion-dilaton fields.
A static string solution, as for instance that of the fundamental
string \cite{dabharv}
is not self-dual (as one may see for instance by the set
of supersymmetries it preserves).

The D-branes of ordinary 10d string theory
carry charges under the antisymmetric tensor fields
of the R-R sector.  In the N=2 string these fields are gauge
equivalent to NS-NS fields by spectral flow in the N=2 U(1); 
effectively, there are no R-R gauge fields, and the D-branes
serve as sources for the NS sector fields (as one sees 
by inserting closed string vertex operators on the disk).
There appear to be no gauge charges 
carried by N=2 string D-branes, apart from those of the self-dual
gravitational field.

\acknowledgements
It is my great pleasure to thank
Martin O'Loughlin and especially
David Kutasov 
for the intense collaboration which led to the
ideas presented here.
Thanks also to 
Tom Banks,
Jeff Harvey,
Chris Hull,
Greg Moore,
and
Stephen Shenker
for fruitful discussions.
The author is supported in part by DOE grant DE-FG02-90ER-40560.

\end{document}